\documentclass[]{article}
\usepackage[UKenglish]{isodate}% http://ctan.org/pkg/isodate
\usepackage[utf8]{inputenc}
\usepackage[english]{babel}

\usepackage{multirow}

\usepackage{xcolor}

\usepackage[margin=2cm]{geometry}

\usepackage{amsmath}
\usepackage{graphicx}
\usepackage{caption}
\usepackage{rotating}

\usepackage[xindy]{glossaries}

\newcommand{\Genom}{G\textsuperscript{en}oM}
\newcommand{\Uppaal}{\textsc{Uppaal}}
\newcommand{\Gwen}{\textsc{GWENDOLEN}}
\newcommand{\Ethan}{\textsc{ETHAN}}
\newcommand{\Rosrv}{\textsf{ROSRV}}
\newcommand{\Rvbip}{RV-BIP}
\newcommand{\Prism}{PRISM}
\newcommand{\Klaim}{\textsc{Klaim}}
\newcommand{\StoKlaim}{\textsc{Sto}\Klaim}
\newcommand{\NuSMV}{\texttt{NuSMV}}

\newcommand{\Mucalc}{$\mu$-calculus}

\newacronym{ltl}{LTL}{Linear-time Temporal Logic}

\newacronym{ctl}{CTL}{Computation Tree Logic}

\newacronym{ptl}{PTL}{Probabilistic Temporal Logic}

\newacronym{pctl}{PCTL}{Probabilistic Computation Tree Logic}

\newacronym{pctl*}{PCTL*}{Probabilistic Computation Tree Logic*}

\newacronym{tctl}{TCTL}{Timed Computation Tree Logic}

\newacronym{stl}{STL}{Signal Temporal Logic}

\newacronym{ltl-x}{LTL-X}{a version of Linear Time Logic without the `next' operator}

\newacronym{mtl}{MTL}{Metric Temporal Logic}

\newacronym{iactlk-x}{IACTLK-X}{a fragment of the temporal-epistemic logic CTLK, without the `next' operator}

\newacronym{fotl}{FOTL}{First-Order Temporal Logic}

\newacronym{bdi}{BDI}{Belief-Desire-Intention}

\newacronym{prs}{PRS}{Procedural Reasoning System}

\newacronym{are}{ARE}{Autonomous Reasoning Engine}

\newacronym{dmars}{dMARS}{distributed Multi-Agent Reasoning System}

\newacronym{fsm}{FSM}{Finite-State Machine}

\newacronym{pfsm}{PFSM}{Probabilistic Finite-State Machine}

\newacronym{apfsm}{APFSM}{\textit{Autonomous} Probabilistic Finite-State Machine}

\newacronym{pha}{PHA}{Probabilistic Hybrid Automata}

\newacronym{fsa}{FSA}{Finite-State Automata}

\newacronym{ta}{TA}{Timed Automata}

\newacronym{gspn}{GSPN}{Generalised Stochastic Petri Net}

\newacronym{mopn}{MOPN}{Marked Ordinary Petri Net}

\newacronym{ba}{BA}{B\"{u}chi Automata}

\newacronym{mdp}{MDP}{Markov Decision Process}

\newacronym{dtmc}{DTMC}{Discrete-Time Markov Chain}

\newacronym{pars}{PARS}{Process Algebra for Robot Schemas}

\newacronym{fsp}{FSP}{Finite State Processes}

\newacronym{piadl}{$\pi$ADL}{$\pi$ calculus combined with ADL}

\newacronym{csp}{CSP}{Communicating Sequential Processes}

\newacronym{ccs}{CCS}{Calculus of Communicating Systems}

\newacronym{wsccs}{WSCCS}{Weighted Synchronous CCS}

\newacronym{csp|b}{CSP$\|$B}{an integrated formalism combining CSP and B}

\newacronym{fdr}{FDR}{the Failures-Divergences Refinement checker}

\newacronym{jpf}{JPF}{Java PathFinder}

\newacronym{ajpf}{AJPF}{Agent Java PathFinder}

\newacronym{mcmas}{MCMAS}{Model Checker for Multi-Agent Systems}

\newacronym{ltlmop}{LTLMoP}{Linear Temporal Logic MissiOn Planning}

\newacronym{adl}{ADL}{Architecture Description Language}

\newacronym{aadl}{AADL}{Architecture Analysis and Design Language}

\newacronym{lts}{LTS}{Labelled-Transition System}

\newacronym{uml}{UML}{Unified Modelling Language}

\newacronym{sysml}{SySML}{Systems Modelling Language}

\newacronym{robotml}{RobotML}{Robot Modelling Language}

\newacronym{dsl}{DSL}{Domain Specific Language}

\newacronym{sct}{SCT}{Supervisory Control Theory}

\newacronym{ide}{IDE}{Integrated Development Environment}

\newacronym{mde}{MDE}{Model-Driven Engineering}

\newacronym{ai}{AI}{Artificial Intelligence}

\newacronym{fdir}{FDIR}{Fault Detection, Isolation and Recovery}

\newacronym{ifm}{iFM}{Integrated Formal Methods}

\newacronym{ros}{ROS}{Robot Operating System}

\newacronym{dl}{\text{\upshape\textsf{d{\kern-0.05em}L}}}{differential dynamic logic}

\newacronym{vm}{VM}{Virtual Machine}

\newacronym{bip}{BIP}{Behaviour Interaction Priority}

\newacronym{fol}{FOL}{First-Order Logic}

\newacronym{owl}{OWL}{Web Ontology Language}

\newacronym{ria}{RIA}{Restore Invariant Approach}

\newacronym{ocl}{OCL}{Object Constraint Language}

\newacronym{sil}{SIL}{Safety Integrity Level}

\newacronym{rule}{RULE}{Robotic Urban-Like Environment}

\newacronym{forms}{FORMS}{FOrmal Reference Model for Self-adaptation}

\makeglossaries

\usepackage{czt}
\usepackage{circus}
\usepackage{hijac}

\usepackage{pdflscape}
\usepackage{multirow}
\usepackage{tabularx}

\usepackage{hyperref}

\makeatletter
\newcommand*{\glsplainhyperlink}[2]{%
  \colorlet{currenttext}{.}% store current text color
  \colorlet{currentlink}{\@linkcolor}% store current link color
  \hypersetup{linkcolor=currenttext}% set link color
  \hyperlink{#1}{#2}%
  \hypersetup{linkcolor=currentlink}% reset to default
}
\let\@glslink\glsplainhyperlink
\makeatother

\usepackage{paralist}

\begin{document}

\title{Formal Specification and Verification of Autonomous Robotic Systems: A Survey\thanks{The first and second authors contributed equally to this work, which was supported by UK Research and Innovation, and {EPSRC}~ Hubs for Robotics and AI in Hazardous Environments: {EP/R026092} (FAIR-SPACE), {EP/R026173} (ORCA), and {EP/R026084} (RAIN). The authors would like to thank Ana Cavalcanti for her early comments on this paper.}}

\author{Matt Luckcuck, \and  Marie Farrell, \and Louise Dennis, \and  Clare Dixon and Michael Fisher \\ \small{Department of Computer Science, University of Liverpool, UK}}
\date{\today}

%\keywords{formal verification, formal specification, autonomous robotics, formal methods}

%\acmJournal{CSUR}

\maketitle

\begin{abstract}
\noindent
Autonomous robotic systems are complex, hybrid, and often safety-critical; this makes their formal specification and verification uniquely challenging. Though commonly used, testing and simulation alone are insufficient to ensure the correctness of, or provide sufficient evidence for the certification of, autonomous robotics.  Formal methods for autonomous robotics has received some attention in the literature, but no resource provides a current overview. This paper systematically surveys the state-of-the-art in formal specification and verification for autonomous robotics. Specially, it identifies and categorises the challenges posed by, the formalisms aimed at, and the formal approaches for the specification and verification of autonomous robotics.
\end{abstract}

\section{Introduction, Methodology and Related Work}
\label{sec:intro}

An \textit{autonomous system} is an artificially intelligent entity that makes decisions in response to input, independent of human interaction. \textit{Robotic systems} are physical entities that interact with the physical world. Thus, we consider an \textit{autonomous robotic system} as a machine that uses \gls{ai}, has a physical presence in and interacts with the real world. They are complex, inherently hybrid, systems, combining both hardware and software; they often require close safety, legal, and ethical consideration. Autonomous robotics are increasingly being used in commonplace-scenarios, such as driverless cars~\cite{Fernandes2017}, pilotless aircraft~\cite{webster2011formal}, and domestic assistants~\cite{webster_formal_2014,Dixon2014}. 

While for many engineered systems, testing, either through real deployment or via simulation, is deemed sufficient; the unique challenges of autonomous robotics, their dependence on sophisticated software control and decision-making, and their increasing deployment in safety-critical scenarios, require a stronger form of verification. This leads us towards using formal methods, which are mathematically-based techniques for the specification and verification of software systems, to ensure the correctness of, and provide sufficient evidence for the certification of, robotic systems.
 
We contribute an overview and analysis of the state-of-the-art in formal specification and verification of autonomous robotics. 
%We avoid the issues of hardware control, general formal methods approaches, and verification of general artificial intelligence techniques. 
 \S\ref{sec:scope} outlines the scope, research questions and search criteria for our survey. 
\S\ref{sec:relatedWork} describes related work concerning formal methods for robotics and differentiates them from our work. We recognise the important role that middleware architectures and, non- and semi-formal techniques have in the development of reliable robotics and we briefly summarise some of these techniques in \S\ref{sec:nonFormal}. %, to provide the reader with the context of robotic systems development. 
The specification and verification challenges raised by autonomous robotic systems are discussed next: \S\ref{sec:extChallenges} describes the challenges of their context (the \textit{external} challenges) and \S\ref{sec:intChallenges} describes the challenges of their organisation (the \textit{internal} challenges). \S\ref{sec:formalisms} discusses the formalisms used in the literature for specification and verification of autonomous robotics. \S\ref{sec:approaches} characterises the approaches to formal specification and verification of autonomous robotics found in the literature. Finally, \S\ref{sec:summaryAndObs} discusses the results of our survey and presents our observations on the future directions of formal methods for autonomous robotic systems.

\begin{sidewaystable}
%\begin{landscape}
%\begin{table}
%   \vspace{-\marginparsep}
%    \vspace{-\marginparwidth}

\scalebox{0.55}{
\begin{tabular}{|c|p{24cm}|p{15cm}|}  
\hline
\textbf{Ref} & \textbf{Formalisms/Tools Employed}  & \textbf{Case Study or Example Used}\\\hline
\cite{webster_formal_2014} & SPIN, Brahms modelling language (BrahmsToPromela translator), LTL. & Care-O-Bot robotic assistant in a Robot House Environment.\\ \hline

\cite{fisher_verifying_2013} & \textsc{Gwendolen}, AJPF, MCAPL, LTL. & RoboCup rescue scenario, autonomous satellite scenario, autonomous pilotless aircraft scenario.\\ \hline

\cite{kouskoulas2013certifying} & Quantified differential dynamic logic (Qd$\mathcal{L}$) and proof using KeYmaeraD. & Formally verified control algorithms for surgical robots.\\\hline

\cite{althoff_safety_2010} & Reachability analysis. & A wrong-way driver threatens two autonomously driving vehicles on a road with three lanes. \\ \hline

\cite{Konur2010} & PCTL models of Probabilistic State Machines, \Prism{}. & The foraging (swarm) robot scenario~\cite{Liu2007}. \\ \hline

\cite{obrien_automatic_2014} & Process Algebra for Robotic Systems (PARS) and MissionLab. & Search for a Biohazard . \\ \hline

\cite{Weyns2010} & Z model for self-adaptive systems, CZT. & MAPE-K reference model~\cite{Kephart2003} and Layered self-adaptive robotics software~\cite{Edwards2009}. \\ \hline 

\cite{mitsch2013provably} & KeYmaera, dynamic differential logic for hybrid systems. & Robotic ground vehicle navigation. \\ \hline

\cite{webster2011formal} & \textsc{Gwendolen}, SPIN, AJPF, LTL. & Model checking for certification of autonomous pilotless aircraft systems.\\ \hline
%10
\cite{basu2008incremental} & BIP (Behaviour-Interaction-Priority) component framework, Evaluator model-checker (temporal logic properties) and monitors. & DALA (iRobot ATRV). \\ \hline

 \cite{chen2012formal} & Theoretical work inspired by LTL and model-checking, computational framework where task specifications are regular expressions. Finite State Automata. & Car-like robots operating in an urban-like environment. \\ \hline
 
 \cite{proetzsch2007formal} & Quartz (synchronous language),model-checking LTL/CTL specifications using Beryl model-checker in the Averest verification framework. & Behaviour-based control network of mobile outdoor robot RAVON. \\ \hline
 
 \cite{chkouri2008translating} & A metamodel-based translation from AADL to BIP. Aldebaran model-checker (in BIP) for deadlock detection and observers for monitoring safety properties. & Flight computer. \\ \hline
 
 \cite{pallottino2007decentralized} & Probabilistic verification using Monte Carlo approach (no tool support, theorems and proofs in paper). & Collision avoidance management of a large and changing number of autonomous vehicles.\\ \hline
 
 \cite{dantam2012robust} & SPIN to verify \textit{Ach} (novel interprocess (IPC) communication mechanism and library). & Implementation of \textit{Ach} on Golem Kang humanoid robot. \\\hline
 
 \cite{iglesia2015mape} & Timed Automata and Timed CTL (TCTL) to verify MAPE-K templates for the development of self adaptive systems, and \Uppaal{}. & Traffic monitoring, outdoor learning robot, transportation and, storytelling robot. \\ \hline
 
 \cite{molnar2009system} & Hybrid system modelling using natural language programming (sEnglish) which is translated to Stateflow and verified using MCMAS. & Autonomous underwater vehicles.\\\hline
 
 \cite{bordini2009formal} &(Extended Abstract) Brahms modelling framework translated into Jason agent code and model-checking temporal logic properties. & Astronaut-robot collaboration.\\\hline
 
 \cite{hess2014formal} & Systems of differential equations and manoeuvrer automata generated using reachability analysis.& Collision avoidance of road vehicles.\\\hline
 %20
 \cite{raman2011analyzing} & LTL, extension to LTLMoP toolkit for robot mission planning. & Fire fighting scenario. \\ \hline
 
 \cite{bensalem2011formal} & BIP framework, DFinder model-checker. & DALA autonomous rover. \\\hline
 
 \cite{risler2008formal} & Extensible Agent Behaviour Specification Language (XABSL) - language for describing state machines. & Robot Soccer. \\ \hline
 
 \cite{proetzsch2007formal} &Synchronous language Quartz and verification using model-checking techniques of the Averest verification framework (Beryl). & Mobile outdoor robot RAVON.\\\hline
 
\cite{iftikhar2012case} & \gls{ta} for modelling and \gls{tctl} for safety and correctness properties for an adaptive autonomous system. & Distributed network of traffic monitoring cameras that reorganise to coordinate data transmission. \\ \hline

\cite{falcone2011runtime} & \Rvbip{}, a tool that produces runtime monitors for robots the BIP framework from formal descriptions of the system and monitor. & Autonomous rover. \\ \hline

\cite{iftikhar2014activforms} & \gls{ta} for modelling different modes of operation, and a module that enables them to be swapped during their execution on a \gls{vm}. & Turtlebot robots navigating a map, representing a goods-transportation system. \\ \hline

\cite{winfield2008modelling} &  \gls{pfsm}, transition probabilities are estimated and then validated by simulation in Player/Stage. & Swarm navigating with the alpha algorithm. \\ \hline
 
\cite{costelha2007modelling} & Layered modelling of a robot's task plans using Petri Nets. & Robot Soccer. \\ \hline

\cite{weyns2012forms} & Reference model for building verifiable self-adaptive systems (FORMS). & Traffic monitoring system.\\\hline

\cite{celaya2007modeling}  & Modelling an abstract architecture for \textit{reactive} multi-agent systems, which are then analysed for deadlock freedom. & Small example with two agents cooperating to move objects. \\ \hline
%30
\cite{ziparo2008petri}  &  Modelling plans as Petri nets with an additional set of Goal Markings for single- and multi-robot systems. & Multi-robot foraging and object passing. \\ \hline

\cite{massink2013use} & Bio-PEPA process algebra to model the Marco and Micro level behaviour of robot swarms. These are analysed using the Bio-PEPA Tool Suite and PRISM. & Swarm foraging, where a team of three robots is needed to carry each object. \\ \hline

\cite{kloetzer2011multi} & LTL-X (LTL without ``next" operator) and B\"uchi Automata  to give a robot control and communications strategy with reduced `stop-and-wait' synchronisations. & A team of unicycle robots with distributed control. \\ \hline

\cite{liang2009software} & Specifies robot behaviour in Z, then links Z animator Jaza to the Java debugger \textbf{\textit{jdb}} to monitor the running program without alterations or additions. & Robot assembly system, for building robots for the NASA ANTS project~\footnote{\url{https://attic.gsfc.nasa.gov/ants/} }.  \\ \hline

\cite{akhlaki2007methodological} & Derives a specification of a real-time system (in CSP+T) from a UML-RT Model for design-time schedulability and dependability analysis. & Two-armed industrial robot and mechanical press. \\ \hline

\cite{gjondrekaj2012towards} & Methodology for verifying three cooperative robots with distributed control using the formal language \textsc{Klaim},  and its extension for stochastic analysis. & Three cooperative robots transporting an object to a goal area. \\ \hline

\cite{alsafi2010ontology} & Ontology-based (\gls{owl}) reconfiguration agent in a manufacturing environment. & Four machines: (1) processing, (2) handling, (3) Simple Conveyor Chain, and (4) filling station.\\\hline

\cite{tenorth2009knowrob} & KnowRob ontology which is based on description logic, \gls{owl}.  & Household robot setting a table. \\\hline

\cite{PRESTES20131193} & IEEE-RAS Ontology for robotics and automation. & N/A \\\hline

\cite{karaman2009sampling} & Generating robot motion plans that satisfy properties in a deterministic subset of \Mucalc{}. &  N/A\\ \hline
%40
\cite{kress2011correct} & Generating robot behaviour automata that satisfy LTL specifications, uses \gls{ltlmop} and TuLiP. & Driverless car in an urban environment. \\ \hline

\cite{filippidis2012decentralized} & Distributed LTL specifications over multiple agents, exchanged when the robots physically meet and model checked for mutual satisfiability. & N/A \\ \hline

\cite{talamadupula2014coordination} & \gls{fol} to formalise agent beliefs and intentions, and uses these to predict the plan of other agents for cooperation. & Willow Garage PR2 Robot, using \gls{ros} in a human-robot team scenario. \\ \hline

\cite{dylla2008approaching} & Uses Readylog to formalise the basic concepts of football theory, in order to formally describe football tactics. & RoboCup football robots. \\ \hline

\cite{bonfe2012towards} & UML, Alloy state machines, KodKod SAT solver, \gls{ltl} and \gls{fol}. & Surgical operations.\\ \hline

\cite{kress2007s} & Fragment of temporal logic called General Reactivity (1). & Search and rescue style missions. \\\hline

\cite{basu2011rigorous} & \gls{bip} framework. & DALA iROBOT ATRV.\\\hline

\cite{bensalem2009designing} & \gls{bip} and LAAS frameworks. & DALA iROBOT ATRV.\\\hline

\cite{guo2013revising} & \gls{ba} modelling the environment and \gls{ltl} task specifications. & A surveillance robot. \\ \hline

\cite{correll2007modeling} & Markov chains to predict swarm aggregation. & Cockroach aggregation. \\ \hline
%50
\cite{hilaire2007formal} & Integration of Object-Z and Statecharts, SAL model checker, and STeP theorem prover. & N/A\\ \hline

\cite{whittle2009relax} & Requirements language for dynamically adaptive systems (RELAX) whose semantics is defined in terms of temporal fuzzy logic. & Adaptive assisted living smart home.\\ \hline

\cite{gudemann2008specification} & \gls{ria}, Microsoft Robotics Studio simulator, JADEX multi-agent framework, Alloy, class diagrams and OCL.  & Adaptive production cell.  \\ \hline

\cite{Dennis2016a} & \gls{ltl} and \gls{ajpf}. & Three pilotless civilian aircraft examples: brake failure, eratic intruder aircraft, and low fuel. \\ \hline

\cite{Dennis2015} & \gls{ltl} and \gls{ajpf}. & Ethical scenario of a robot trying to stop humans from falling down a hole. \\ \hline

\cite{huang2014rosrv} & \gls{ltl}. & LandShark rover robot. \\ \hline

\cite{abdellatif2012rigorous} & \gls{bip} models to generate \Genom{} controllers, and D-Finder tool. & Dala rover navigating to a target position, taking a picture, and returning to the initial position. \\ \hline

\cite{konur2012analysing} & \glspl{dtmc} to model the foraging algorithm, \gls{pctl} for safety properties, \Prism{}. & Foraging robot swarm. \\ \hline

\cite{brambilla2012property} & \glspl{dtmc} to model the aggregation algorithm, \gls{pctl*} specifications, and \Prism{} for checking. & Aggregation in robot swarm. \\ \hline

\cite{dixon2011towards} & \glspl{fsm}, \gls{ptl} for specification, checked in \NuSMV{}. & Alpha algorithm for robot swarms. \\ \hline
%60
\cite{dixon2012towards} & \glspl{fsm}, \gls{ptl} for specification, checked in \NuSMV{}. Also a NetLogo simulation. & Alpha algorithm for robot swarms. \\ \hline

\cite{chen2013formal} & An update of \cite{chen2012formal} that uses JFLAP and MatLab. & Robotic urban-like environment (RULE). \\\hline

\cite{webster_toward_2016} & BrahmsToPromela and SPIN model-checker. & Care-O-Bot. \\\hline
%63
\end{tabular}}
\caption{These are the papers that were obtained using the methodology outlined in \S \ref{sec:scope} that were deemed to be in scope.}
\glsresetall
\label{table:littable}

\end{sidewaystable}

\subsection{Methodology}
\label{sec:scope}
In this section, we outline the methodology that was followed when conducting this review. First we discuss the factors that delimit the scope of our survey, then we identify the research questions that we analysed, and outline the search criteria that were followed. 

\paragraph{\textbf{Scope:}}
Our primary concern is the formal specification and verification of autonomous robotic systems. This hides a wealth of detail and often conflicting definitions. Therefore, we delimit the scope of our survey as follows:
\begin{compactitem}
\item We target systems that (eventually) have some physical effect on the world, not purely software-based systems. In this sense, some of the work that we examine falls within, though does not fully cover, the realm of \emph{cyber-physical systems} (which are typically tackled by hybrid formalisms);

\item We cover not only systems that affect humans but also those that can be controlled by humans, though we do not model human behaviour in detail as this is beyond the current capabilities of formal methods.

\item While some robotic systems may, as above, be controlled remotely by a human, we allow for the full range of autonomy, from human controlled, to adaptive (i.e. driven by environmental interactions) and on to fully autonomous systems (i.e. that can choose to ignore environmental stimuli and make their own decisions).

\item We address a range of formal properties, including safety, security, reliability, efficiency, etc., primarily concerning the functional behaviour of autonomous robotic systems software. In particular, we do not consider mechanical, materials, or physics issues since our focus is on the software that is controlling these systems. 

\item By `formalisms' we mean mathematically-based techniques for software and systems development. We do not cover techniques based on, for example, systems of  differential equations except to mention them as part of hybrid approaches.

\end{compactitem}

\paragraph{\textbf{Research Questions:}} Our objectives in this paper are to enumerate and analyse the formal methods used for the formal specification and verification of autonomous robotic systems. To this end, we seek to answer the following three research questions: 
\begin{compactdesc}
\item[RQ1:] What are the challenges when formally specifying and verifying the behaviour of (autonomous) robotic systems?
\item[RQ2:] What are the current formalisms, tools, and approaches used when addressing the answer to \textbf{RQ1}?
\item[RQ3:] What are the current limitations of the answers to \textbf{RQ2} and are there developing solutions aiming to address them?
\end{compactdesc}
There are, of course, many other questions that could be posed but we begin with the above. Indeed, these are enough to expose a wide variety of research in the core areas. We answer \textbf{RQ1} in \S\ref{sec:challenges} and \S\ref{sec:autochall} where we identify the current \textit{external} and \textit{internal} challenges, respectively, to formally specifying and verifying autonomous robotic systems that we have derived from our examination of the literature. We provide a thorough analysis of our findings in \S\ref{sec:formalisms} and \S\ref{sec:approaches}, where we examine potential answers to \textbf{RQ2} and \textbf{RQ3}. \S\ref{sec:discussion} provides a thorough discussion of how we answer these research questions and our observations of the literature. 

\paragraph{\textbf{Search Criteria:}}
Our search queries were \textit{formal modelling of (autonomous) robotic systems}, \textit{formal specification of (autonomous) robotic systems}, and \textit{formal verification of (autonomous) robotic systems}.
Our search travelled 5 pages deep of Google Scholar (on 21/05/2018) and discounted any results that were obviously out of scope. In total, we surveyed 156 papers of which 63 were deemed to be in scope and these are  described in Table \ref{table:littable}. We restricted our search to papers that were published in the last ten years (2007 -- 2018). Note that we include those that were published up until the date the search was carried out in 2018. To contextualise these results we did some `snowballing' from the core set of 63 papers and these will be mentioned throughout this survey, however, our analysis specifically focuses on this core set in order to preserve the integrity of our results.

We have broadly followed the methodology in \cite{kitchenham2007guidelines} by delineating research questions, scope and search criteria. However, unlike the usual systematic review, we have employed `snowballing' to provide a more rounded view of the topic and to minimise omissions. Generally, systematic reviews provide a detailed enumeration of the tools or approaches used in a specific domain. They do not normally, however, describe or discuss these tools or approaches in great detail. We have crafted this survey to not only summarise and enumerate the current state-of-the-art, but also to educate the reader and provide a point of reference for researchers in this area.

\subsection{Related Work}
\label{sec:relatedWork}

To our knowledge, there is no survey of the recent literature for formal specification and verification of autonomous robotic systems. This section differentiates related work from ours.

Robotic systems often have safety-critical consequences. A survey on safety-critical robotics~\cite{Guiochet2017} identified seven focus areas for the development of robots that can safely work alongside humans and other robots, in unstructured environments. These areas are: (1) modelling and simulation of the physical environment to enable better safety analysis, (2) formal verification of robot systems, (3) controllers that are correct-by-construction, (4) identification and monitoring of hazardous situations, (5) models of human-robot interaction, (6) online safety monitoring that adapts to the robot's context, and (7) certification evidence. In contrast, our survey focusses on the application of formal methods to any type of autonomous robotic system.

Nordmann et al.~\cite{Nordmann2016} present a detailed survey of Domain Specific Modelling Languages for robotics. Their study focusses on non-formal modelling, but was influential to the depth and organisation of our survey. They analyse languages by both the applicable domain and development phase. The use of model-driven engineering approaches in building robotic systems seems too prevalent to ignore, as we discuss in \S\ref{sec:nonFormal}.

In applying formal verification methods to programming languages for autonomous software, Simmons et al.~\cite{simmons_towards_2000} identified three challenges for the automatic verification of autonomous systems: (1) ``special purpose languages use special-purpose interpreters or compilers'' that must be verified, (2) ``application-specific programs written in these languages need to be verified for internal correctness'' (safety, liveness, etc.) and (3) ``programs need to be verified for external correctness'' (how they interact with the overall software system). 

Weyns et al.~\cite{Weyns2012} present a survey of formal methods for self-adaptive systems. While robotic systems may be required to adapt themselves, this behaviour is not required of all robotic systems and so it is somewhat tangential to our survey. We discuss the specific challenges of formally specifying and verifying self-adaptive robotic systems in \S\ref{sec:reconfig}.

Rouff et al.~\cite{rouff2004formal, Rouff2004} survey formal methods for intelligent swarms. As we discuss in \S\ref{sec:swarms}, the challenges posed by robot swarms are an extension of those of single autonomous robotic systems. They compare a wide range of formalisms, covering process algebras, model-based notations, various logics, amongst others. They conclude that no single formalism is suitable to the range of challenges posed by intelligent robot swarms. Their focus is, again, narrower than ours and \S\ref{sec:swarms} offers a more recent perspective on formal methods for robot swarms.

\section{General Software Engineering Techniques for Robotic Systems}
\label{sec:nonFormal}

The complex task of engineering a reliable robotic system is often addressed using a variety of software engineering techniques. Though the focus of this survey is formal approaches to specifying and verifying robotic systems, this section introduces some of the less formal approaches; particularly where they are potentially useful inputs to, or could be integrated with, a formal technique. We intend this section to provide the reader with a more rounded and complete view of robotic system engineering, beginning with the commonly used middleware architectures.

\paragraph{\textbf{Middleware Architectures:}} %%Architectures
 Architectures for developing robotic systems generally adopt the paradigm of communicating \textit{components}/\textit{nodes}. %These simplify the engineering process and encourage the reuse of existing software. 
 Among the most popular in the literature are: ROS~\cite{quigley2009ros}, OPRoS~\cite{Jang2010}, OpenRTM~\cite{Ando2008}, Orocos~\cite{Schlegel1999}, and \Genom{}~\cite{Fleury1997}. Each of these supports engineering a robotic system in a slightly different way but most share a common set of component-based, modular concepts~\cite{Shakhimardanov2010} % The concepts common to Orocos, ROS, \Genom{}, and OpenRTM are identified in~\cite{Shakhimardanov2010}; 
 to facilitate reuse of modules between different frameworks.
Supporting the integration of middleware frameworks, \cite{Ahn2017} presents a robotic software manager with a high-level API that bridges the gaps between ROS, OPRoS, and OpenRTM.% Their top-down approach allows a single system to be engineered from components developed using multiple middleware frameworks. These two abstractions could be the basis of a formal technique for describing component-based systems. 
\medskip

\noindent It is often assumed that a middleware is sound, and this is key to trusting the robotic system~\cite{Foughali2016}. However, given the heterogeneity of the systems that can be produced using such architectures and their parametrisable nature, guaranteeing correct robot behaviour is challenging~\cite{Halder2017}. Thus, we have identified the following categories of non- and semi-formal methods.

\paragraph{\textbf{Testing and Simulation:}} 
Both testing and simulation are useful when developing robotic systems. Particularly simulation, which enables the examination of the system's behaviour in statistically unlikely situations. However, they are time-consuming, and examine part of the program's state space, so they cannot be used to reason reliably about properties of the whole program. Moreover, field tests are potentially dangerous to life and the robot hardware Examples include the following uses of field tests~\cite{Konur2010, Foughali2016, Halder2017, Moarref2017} and/or simulations~\cite{Konur2010, alami2007provably}.

\paragraph{\textbf{\glspl{dsl}:}} %%DSLs/Programming are used to describe specific sections of a robotic system.
  An extensive literature survey identified 137 state-of-the-art \glspl{dsl} for robotics~\cite{Nordmann2016}, the majority of which were for capturing robotics architectures and for use within mature subdomains (e.g. robot motion). Conversely, fewer \glspl{dsl} were aimed at emerging subdomains (e.g. force control). The majority of surveyed \glspl{dsl} were designed to describe the basic components of the system at the early stages of development, but very few were suited to application at runtime. The identified \glspl{dsl} were often defined using Ecore~\cite{steinberg2008emf} meta-models. %, which could be a useful point of convergence to encourage the sharing of meta-models. 
 %This seems like a useful avenue of future work for integrating formal methods more closely with existing engineering practices. 
 Related work on providing a robotics-specific \gls{ide}~\cite{Stampfer2016} aims to integrate the various phases of robotics software development %. They integrate a variety of \glspl{dsl} into an \gls{ide} that 
 by guiding the user through a standardised development process. The Declarative Robot Safety language (DeRoS)~\cite{Adam2016} is a DSL for describing a robot's safety rules and corresponding corrective actions. Usefully, this provides automatic code generation to integrate these rules with runtime monitoring by generating a ROS safety monitoring node. 

\paragraph{\textbf{Graphical Notations:}}%% 
Graphical Notations are present and examined throughout the literature. They usually aim to provide an easy to use mechanism for designing robotic software that enables communication between engineers from the various distinct disciplines involved in robotics \cite{Harel1987, Wachter2016, Merz2006}.
Statecharts~\cite{Harel1987} form the basis for several of these graphical notations, such as ArmarX statecharts~\cite{Wachter2016} which are executable as C++ programs.  % For example, ArmarX statecharts~\cite{Wachter2016} provide a generic method for describing robotic systems. Each state has four different phases: \textit{onEnter}, \textit{running}, \textit{onBreak}, and \textit{onExit}. Individual phases are linked to a user-defined function enabling developers to run their own code during different phases of a state. Transitions between states facilitate the transfer of data and ArmarX Statecharts are executable, as C++ programs.
A similar executable notation is \textit{restricted Finite State Machine} (rFSM) Statecharts~\cite{Merz2006}, which represent a subset of UML 2 and Statecharts~\cite{Harel1987}. %Again, states have actions that occur at different phases of the state: \textit{entry}, \textit{exit}, and the in-state activity \textit{do}. States perform an action in response to an event. Transitions between states can have a boolean guard and an action that occurs during the transition. 
Other notations include RoboFlow~\cite{Alexandrova2015} which adopts the style of flow charts. %, boxes represent procedures and directed arrows between them indicate data flow. Diamonds represent a choice, with one input and multiple outputs. %No control flow constructs (such as \textit{while} or \textit{repeat}) are present in this notation.
 RoboFlow is specifically designed for engineering robot movement and manipulation tasks.

\paragraph{\textbf{MDE/XML:}} Numerous \gls{mde} approaches use the Eclipse Modelling Framework's Ecore meta-meta-model \cite{fleurey2007generic}. The work in~\cite{Hua2016} presents a mapping from the XML-based AutomationML into ROS program code. The BRICS Component Model~\cite{Bruyninckx2013} translates high-level models into platform-specific component-models for Orocos and ROS. Another model-driven approach is the MontiArcAutomaton architecture description language~\cite{Ringert2015} which translates models into program code, mostly for ROS~\cite{Adam2017}. Performance Level Profiles~\cite{Brafman2016} is an XML-based language for describing the expected properties of functional modules. They provide tools for the automatic generation of code for runtime monitoring of described properties. They again target the ROS framework, but the output from their tool is flexible and not tied to ROS. RADAR is a novel procedural language, with a Python implementation, for describing robot event responses~\cite{Biggs2006}. 

\paragraph{\textbf{Discussion:}}
Middleware frameworks provide a flexible standard interface to the robotic hardware and the above non- and semi-formal approaches aim to adapt the parts of software engineering practice that are relevant to the robotics domain. \glspl{dsl} abstractly describe properties about specific parts of the system, often improving the understanding or interoperability of the system. Graphical notations aid in communication and visualisation. However, more formal languages provide unambiguous semantics and enable reasoning about the whole program rather than only a particular set of program behaviours. Next, we describe the \textit{external} challenges faced when applying formal methods to robotic systems. These challenges are external to the robotic system and are independent of the type of robotic system being developed.

\section{External Challenges of Autonomous Robotic Systems}
\label{sec:challenges}
\label{sec:extChallenges}

This section discusses the inherent challenges of formally specifying and verifying any autonomous robotic system. We categorise them as \textit{external} challenges because they are independent of the robotic system's internal design.

\S\ref{sec:environment} discusses the  challenge of modelling a robotic system's physical environment and verifying its behaviour within that environment. This process is of paramount importance because real-world interactions can significantly impact safety.

\S\ref{sec:trustAndCertEvidence} discusses the challenge of providing sufficient evidence for both certification and public trust. For some robotic systems, the domain in which they are deployed has a certification body (for example, in the nuclear industry) that describes the evidence required to certify the system for use. Other robotic systems are being used in domains where there is no certification body (domestic assistants, for example) so evidence must be provided to gain trust in their safety. Both situations are challenging because of the technical nature of both robotic systems and formal verification.

\subsection{Modelling the Physical Environment}
\label{sec:environment}

As mentioned in \S\ref{sec:intro}, we consider an autonomous robot as a machine that implements \gls{ai} techniques, has a physical presence, and interacts with the world. Thus, one of the most prominent challenges in verifying robotic systems is its interaction with an unstructured environment. This is further complicated by differing sensor accuracy or motor performance, or components that are degraded or damaged. 

Interactions between a robot and its physical environment can have a major influence on its behaviour, because the robot may be reacting to environmental conditions not considered at design-time. Indeed, the behaviour of adaptive systems exhibit is directly driven by environmental interactions. While it is accepted that formally modelling a robotic system \textit{within} its physical environment is important~\cite{walter_experiences_2010, fisher_verifying_2013}, it has had limited attention in the literature. We believe that this is due to the sheer complexity of this task: aside from the difficulty of modelling the real world, a robot only has partial knowledge of its surroundings. Sensing limitations caused by sensor blind-spots and interference between sensors add extra complexity to the modelling process~\cite{bouraine2012provably, phan2015collision, hess2014formal}. 

To address the challenge of combining discrete computations and a continuous environment, a robotic system is typically separated into several layers~\cite{alami_architecture_1998,fisher_verifying_2013}. At the bottom, the functional layer consists of control software for the robot's hardware. Then, the intermediate layer generally utilises a middleware framework (such as ROS~\cite{quigley2009ros} or \Genom{}~\cite{Fleury1997}) that provides an interface to the hardware components. The upper layer contains the decision making components of the robotic system, which capture its autonomous behaviour. 

Some previous work has focussed on a robot's decision making, ignoring its environment~\cite{Lopes2016, Konur2010}. Others assume that the environment is static and known, prior to the robot's deployment~\cite{Moarref2017, webster_formal_2014, Gainer2017}, which is often neither possible nor feasible \cite{hess2014formal}. For example, the environment may contain both fixed and mobile objects whose future behaviour is unknown \cite{bouraine2012provably} or the robot's goal may be to map the environment, so the layout is unknown. Hybrid architectures often take a third approach; to abstract away from the environment and assume that a component has the ability to provide predicates that represent the continuous sensor data about the robot's environment~\cite{AbstrHybSys08, Book:VerificationHybridSys09, fisher_verifying_2013, Dennis2015, Dennis2016a}. While this encapsulates the high-level control, insulating it from the environment, the implicit assumption of a static, known environment can often remain.

Using models of the environment that are static and assume prior knowledge may limit their effectiveness. However, Sotiropoulos et al.~\cite{Sotiropoulos2017} examine the ability of low-fidelity environmental simulations to reproduce bugs that were found during field tests of the same robot. Of the 33 bugs that occurred during the field test, only one could not be reproduced in the low-fidelity simulation. Perhaps similar results might be obtained with low-fidelity formal models of a robot's environment.

Probabilistic models are a popular approach to capturing dynamic and uncertain environments. In~\cite{Morse2016}, a PRISM model captures the environment of a domestic assistant robot. Non-robot actors in the environment are specified using probabilistic behaviours, to represent the uncertainty about their behaviour. The robot model is also written in PRISM, so that it can be reasoned about within this probabilistic environment. This is a useful step, that accepts the changing nature of the environment. However, for the model-checker to explore outcomes based on a certain behaviour, its probability must be encoded into the environment model. This still leaves the possibility of unforeseen situations having a detrimental effect on the robot's behaviour.

Similarly, Hoffmann et al.~\cite{Hoffmann2016} formalise a pilotless aircraft and its physical environment using an extension of \glspl{pfsm}. They use \Prism{} to verify properties about their model, with the pilotless aircraft tasked with foraging for objects, which it must return to a drop-off location. Their approach involves running small-scale simulations of the pilotless aircraft in its environment to determine the probabilities and timing values for their formal model. 
In contrast, Costelha and Lima~\cite{costelha2007modelling} divide their models into layers and try to obtain a model of the robot's environment that is as realistic as possible. Their work uses Petri Nets with the environment model using untimed direct transitions and other layers in the models using stochastic Petri Nets.

Modelling the environment is particularly relevant for navigation, and tackling collision avoidance and safe robot navigation~\cite{phan2015collision, bouraine2012provably, mitsch2013provably, proetzsch2007formal} often feature in the literature. Although several navigation algorithms exist, some cannot be used in safety-critical scenarios because they have not been verified~\cite{phan2015collision}. The Simplex architecture~\cite{seto1998simplex} provides a gateway to using these unverified algorithms with the concept of advanced controllers (AC). ACs are used alongside a verified baseline controller (BC) and a decision module, which chooses which controller is active. The decision module monitors the system and if it determines that the AC will violate one of the safety properties, then the BC takes control. 
Other work employs reachability analysis to generate a manoeuver automaton for collision avoidance of road vehicles~\cite{hess2014formal}. Here, differential equations model the continuous behaviour of the system. Runtime fault monitoring is useful for catching irregular behaviours in running systems but it does not ensure safety in all situations.

Mitsch et al.~\cite{mitsch2013provably} use \gls{dl}~\cite{Platzer2007}, designed for hybrid systems, to describe the discrete and continuous navigation behaviour of a ground robot. Their approach uses hybrid programs for modelling a robot that follows the dynamic window algorithm and for modelling the behaviour of moving objects in the environment. Using the hybrid theorem prover KeYmaera, Mitsch et al. verify that the robot will not collide with, and maintains a sufficient distance from, stationary and moving obstacles. In proving these safety properties, 85\% of the proof steps were carried out automatically. 
 
In further work, Mitsch et al.~\cite{Mitsch2017} verify a safety property that makes less conservative driving assumptions, allowing for imperfect sensors; and add liveness proofs, to guarantee progress. They extend their approach to add runtime monitors that can be automatically synthesised from their hybrid models. They note that all models deviate from reality, so their runtime monitors complement the offline proofs by checking for mismatches between the verified model and the real world.

Formal and non-formal models of the real world are prone to the problem of the \textit{reality gap}, where models produced are never close enough to the real world to ensure successful transfer of their results~\cite{walter_experiences_2010,fisher_verifying_2013}. This is especially problematic when real-world interactions can impact safety. Bridging the gap between discrete models of behaviour and the continuous nature of the real-world in a way that allows strong verification is often intractable~\cite{Desai2017}. Moreover, in a multi-robot setting one robot can be part of another's environment \cite{kress2007s}. There is also a trade-off between ensuring that the system is safe and ensuring that it is not so restrictive that it is unusable in practice \cite{walter_experiences_2010}.

\subsection{Evidence for Certification and Trust}
\label{sec:trustAndCertEvidence}

Robotic systems are frequently deployed in safety-critical domains, such as nuclear or aerospace. These systems require certification, and each domain usually has a specific certification body. Ensuring that the engineering techniques used for developing robotic systems are able to provide appropriate certification evidence is, thus, essential. However, autonomous robotic systems development often falls short of providing sufficient evidence, or evidence in the correct form, for these certification bodies~\cite{Foughali2016}.

When certifying a piloted aircraft, the certification bodies assume that it will have a suitably qualified and competent crew in control. When an autonomous system is in control of a pilotless aircraft, this system must also be certified. However, the regulatory requirements for such a pilotless aircraft are still under development~\cite{Webster2014}. Choosing an appropriate formal method for a particular system is no mean feat as there are currently no standard guidelines available to aid developers in choosing the most suitable approach \cite{kossak2016select}. In fact, one of the main difficulties here is in choosing a formalism so that a good compromise is reached between how expressive the formalism is and the complexity of its analysis/synthesis algorithms \cite{belta2007symbolic}. This also presents a challenge for certification bodies who need to be able to determine the reliability of safety-critical robotic systems and the formal methods that were employed during development.  

Autonomous vehicles are usually regulated by the relevant vehicle authority for the territory that they are used in. However, their introduction also requires public trust. The majority of respondents of a recent survey on public attitudes towards driverless cars (UK, USA, and Australia) had both a high expectation of their benefits and a high level of concern over their safety and security~\cite{SchoettleBrandon2014}. 

In contrast to the nuclear and aerospace domains, robots used in domestic or care environments do not have strong certification requirements but must be shown to be safe and trustworthy. This covers both safety and the public perception of safety; notions of usability and reliability; and a perception that the robot will not do anything unexpected, unsafe or unfriendly~\cite{Dixon2014}. This lack of both trust and safety assurances can hamper adoption of robotic systems in wider society~\cite{Gainer2017}, even where they could be extremely useful and potentially improve the quality of human life.

Isabelle/HOL and temporal logic have been used to formalise a subset of traffic rules for vehicle overtaking in Germany~\cite{rizaldi2017formalising}. As we move towards driverless cars a key concern is enumerating how much one robot needs to know about the goals and constraints of other vehicles on the road in order to not be considered to blame for any accident~\cite{mitsch2013provably}. In related work, Webster et al.~\cite{webster2011formal} utilise model-checking to provide certification evidence for an autonomous pilotless aircraft system using the \textsc{Gwendolen} agent language and the \gls{ajpf} approach as an improvement over Promela and SPIN.

The certification process often requires safety cases which provide a structured argument of a system's safety in a certain context, given certain assumptions. %They link claims of safety to the evidence. 
Automating safety case generation is challenging; safety cases must combine elements of the physical design, software design, and the maintenance of both. Denney and Pai~\cite{denney2014automating} have described a methodology for automatically generating safety cases for the Swift pilotless aircraft system using a domain theory and their A\textsc{uto}C\textsc{ert} tool~\cite{denney2008software}, which automates the generation of safety case fragments.

Related to trust is the issue of ethical robotic behaviour. Several approaches to constraining the behaviour of a robotic or autonomous system have focussed on the distinction between ethical and non-ethical choices~\cite{Arkin2009, Arkin2012, Brutzman2013}. Worryingly, they ignore the potential situation where no ethical choice is available, and there seems to be little consideration of this situation in the literature~\cite{Dennis2016a}. Clearly, reasoning about ethical behaviour is at least as difficult with robots as it is with humans.

\subsection{Summary of External Challenges}
\label{sec:issuessummary}

This section identified two main external challenges to the specification and verification of robotic systems in the literature. External challenges come from the operating and design environment of the robotic system. The challenges are (1) modelling the physical environment and (2) providing sufficient (and appropriate) trust and certification evidence. With respect to modelling the physical environment, two approaches appear to dominate the literature, these are to either model or monitor the physical environment. Temporal logics, for example, have been used to model the environment with model-checking used for verification. Specifying a monitor to restrict the robotic system to safe behaviours within its environment reduces the verification burden, as only the runtime monitor needs to be verified. However, comprehensively capturing \emph{all} unsafe situations so that they can be mitigated in advance is extremely difficult. 

Certain tasks, such as navigating an unknown and dynamic environment, are challenging for robotic systems; and a number of navigation algorithms exist in this domain. However, not all can be employed in safety-critical scenarios as they have not been verified~\cite{phan2015collision}. This suggests that there are limitations in the use of current formal methods to verify these algorithms, and leads to hybrid approaches that have high computational complexity. 
Clearly, a recurring challenge is the complexity (both computationally and in terms of usability) of formal models, particularly if we consider the modelling of a complex physical environment.

Formal techniques can be used to provide trust and certification evidence, but it is clear that current techniques are insufficient. This is further complicated by the small number of available guidelines to help developers to identify the most appropriate formal method(s) to specify and verify their system~\cite{kossak2016select}. Regulators are more concerned with the evidence of safety, than with the specific methods used to produce the evidence; therefore, determining suitable and robust formal methods for distinct robotic systems remains an open problem. 

These are the challenges, \textit{external} to the robotic system, that our survey identified. Next, we discuss the \textit{internal} challenges, which stem from how a robotic system is engineered.

\section{Internal Challenges of Autonomous Robotic Systems}
\label{sec:autochall}
\label{sec:intChallenges}

Autonomous robotic systems are increasingly prevalent, and are attractive for removing humans from hazardous environments such as nuclear waste disposal or space exploration \cite{Gao2016}. Our survey identified agent-based approaches as a popular way of providing the autonomy that these systems need; we discuss the challenges that this approach brings to formal specification and verification in \S\ref{sec:agents}. Multi-robot systems are also an active area of research; \S\ref{sec:swarms} and \S\ref{sec:teams} describe the challenges of formally specifying and verifying homogeneous and heterogeneous multi-robot systems, respectively. In hazardous environments, autonomous robotic systems need to be capable of adaptation and reconfiguration in response to changes in the environment, system, or mission; \S\ref{sec:reconfig} discusses the challenges posed by these systems.

\subsection{Agent-Based Systems}
\label{sec:agents}

An \textit{agent} encapsulates the system's decision-making capability into one component. This helps to provide \textit{rational} autonomy (where a system can explain its reasoning) which is crucial for providing evidence to gain public trust or for certification (discussed in \S\ref{sec:trustAndCertEvidence}).  Since the implementation of \gls{prs}~\cite{georgeff1987reactive}, \gls{bdi} agents received a great deal of attention~\cite{rao:95b}. \gls{bdi} systems can use a single or multiple agents, depending on their requirements. The \gls{bdi} model of agency appears to be prevalent in the surveyed literature.

The \gls{dmars} is an implementation of \gls{prs} with the \gls{bdi} model operationalised by plans~\cite{d2004dmars}. A formal Z specification of \gls{dmars} provides a formal foundation for the semantics of \gls{dmars} agents~\cite{d2004dmars}. These agents monitor both the world and their own internal state. They are comprised predominantly of a plan library, functions for intention, event and plan selection, belief domain and goal domain~\cite{d2004dmars}.
  
Webster et al.~\cite{webster2011formal, Webster2014} verify an autonomous pilotless aircraft using a program model-checker. The aircraft is controlled by a \gls{bdi} agent, written in the \Gwen{} language~\cite{Dennis2008}, and verified by the \gls{ajpf} model-checker~\cite{Dennis2012}. The verified agent is plugged into a flight simulator~\cite{Webster2014} to visualise the verified scenarios. The work in~\cite{Kamali2017, Kamali2018} describes a \Gwen{} agent controlling a driverless car in a vehicle platoon. \Gls{ajpf} is used to verify safety properties related to the car joining and leaving a platoon, and for maintaining a safe distance during platooning. Verification of the timing properties is handled by \Uppaal{}. 

Similarly, Choi et al. model a system of nine heterogeneous agents with one representing  the robot's physical environment \cite{choi2015verification}. They used the \gls{mcmas} \cite{lomuscio2009mcmas} to verify the system and illustrated that \gls{mcmas} performed dramatically better than SMV in this setting. 
Molnar and Veres also used \gls{mcmas} to verify autonomous underwater vehicles \cite{molnar2009system}. They use natural language programming (sEnglish) which results in an ontology that is used to represent a labelled transition system. This is then translated into Stateflow and verified with \gls{mcmas}. Interestingly, they represent the human interface as an agent, as does \cite{bordini2009formal}. This links human behaviour and the formal model that they have developed.

Hoffmann et al. characterise the autonomous behaviour of an agent (based on the \gls{bdi} model) as a \gls{pfsm} extended with a weight function to map weights onto actions, which they call an \gls{apfsm} \cite{Hoffmann2016}. Due to the addition of the weight function, an \gls{apfsm} is a discrete-time Markov chain. They use this formalism of agent autonomy to verify properties about a pilotless aircraft on a foraging mission.

Bozzano et al.~\cite{bozzano2011comprehensive} propose the \gls{are} for controlling spacecraft, it provides plan generation, validation, execution, monitoring and fault-detection. It communicates with the ground station, performs autonomous reasoning, receives sensor data, and sends commands to actuators. A prototype \gls{are}, developed using \NuSMV{}, has been used in ESA rover and satellite projects to provide a model-based approach to on-board autonomy. 

Podorozhny et al. \cite{podorozhny2007verification} use Alloy to verify multi-agent negotiations. Their work enables properties of agent coordination and interaction to be checked, and properties of complex data structures (which may be shared between agents). Their work centres on a system where multiple agents cooperate to reach and negotiate to optimise achieving their common goal(s). They model agents' negotiations as \glspl{fsm}, which are then translated into an Alloy specification and verified using Alloy Analyzer. The main challenge was ensuring tractability of the Alloy models.

\subsection{Multi-Robot Systems}
\label{sec:multirobot}

Robotic systems composed of several robots present challenges that must be tackled in addition to those faced when developing single-robot systems. Two broad categories of multi-robot systems exist: \textit{swarms}, where all of the robots exhibit the same behaviour; and \textit{teams}, where the robots may each behave distinctly. In both cases, the behavioural requirements of the system must be considered at both the microscopic level (individual robots) and  at the macroscopic level (whole system). Multi-robot systems can present difficulties for ensuring safety and liveness properties due to the number of concurrent, interacting agents and the changeability of the system's environment~\cite{Akhtar2014}. 

The work in~\cite{gjondrekaj2012towards} presents a methodology for verifying cooperative robots with distributed control. They use the formal language \Klaim{}, which was designed for distributed systems; its stochastic extension, \StoKlaim{}; and its related tool set. They formalise a specification of the robot's behaviour and physical environment, in \Klaim{}; and then add stochastic execution times for actions, in \StoKlaim{}. They validate their modelling approach by comparing their results to those obtained from physics-based simulations of the same system. The case study contains three homogeneous robots, cooperating under distributed control, to transport objects. In this system, the robots have homogeneous behaviour, but are too few in number to be considered a swarm. 

One approach to easing multi-robot system development formally relates the languages used at the macro- and micro-levels of abstraction. To this end, Smith and Li~\cite{smith2014maze} present MAZE, an extension of Object-Z for multi-agent systems, which includes Back's action refinement. They describe a top-down refinement-based development process and several shorthand notations to improve robot specification at the micro-level.

Next, we discuss the specific challenges faced when developing homogeneous robot systems, specifically swarms, (in \S\ref{sec:swarms}) and heterogeneous robot systems, specifically teams, (in \S\ref{sec:teams}).

\subsubsection{Swarms}
\label{sec:swarms}

A robot swarm is a decentralised collection of robots that work together~\cite{Beni2004}, often taking inspiration from swarms of ants or bees. Robot swarms generally contain robots that are: adaptive, large in number, not grouped together, relatively incapable or inefficient, and able to send and communicate locally~\cite{Sahin2005}. However, these criteria are not exhaustive, and should simply be used to decide to what degree the term ``swarm robotics'' might apply~\cite{Dorigo2004}.

Formal methods have obvious benefits for specifying the behaviour of swarm robot systems, because field test and simulations only cover a particular instance of the swarm's behaviour~\cite{Konur2010}. However, the large number of robots in a swarm can explode the state space and thus hinder verification using conventional model-checking approaches~\cite{Konur2010}. A further issue is scalability: if you prove a property for a swarm of $n$ robots, does the property hold for $n \pm 1$ robots?

Robot swarms are often used for their resilience. Winfield and Nembrini~\cite{Winfield2006} examined the fault tolerance of robot swarms navigating using the alpha algorithm, and found that robustness arises from the inherent properties of a swarm: simple parallel robots, operating in a fully distributed way, with high redundancy. Of particular interest, they found that the \textit{partial} failure of a robot was more problematic than a robot's total failure. In their example, a robot that was fully functioning, apart from its wheels not turning, anchored the swarm in place. 

Determining if failures will propagate through the swarm, and determining a lower bound on the number of robots that must remain functioning for the swarm to complete its mission is an active area of research~\cite{Kouvaros2017}. Kouvaros and Lomuscio~\cite{Kouvaros2017} present an approach that uses a temporal logic and fault injection to determine the ratio of faulty to functioning robots in a swarm navigating using the alpha algorithm. They specify temporal-epistemic properties and then they find the threshold at which the swarm no longer satisfies these properties. Crucially, their approach is designed to work with swarms where the number of agents is unknown at design-time.

Programs for robot swarms are often developed in an ad-hoc manner, using trial and error. The developers tend to follow a bottom-up approach that involves programming individual robots before combining them to form a swarm which displays a certain set of behaviours~\cite{Brambilla2014,Lopes2016,Moarref2017}. This development process can reduce the effectiveness of formal engineering approaches; the application of formal methods often does not guarantee that the implementation matches the specification, because the programs are built manually from the specification~\cite{Lopes2016}. Winfield et al.~\cite{winfield2008modelling} present a straightforward approach to modelling the macroscopic behaviour of the swarm using a \gls{pfsm}. First, they estimate the transition probabilities in the \gls{pfsm}; then, they validate these probabilities by simulations in Player/Stage\footnote{\url{http://playerstage.sourceforge.net}}.

Markov chains have been used to model the probabilities of members of a robot swarm joining or leaving aggregates of robots~\cite{correll2007modeling}. The probabilities were a function of the size of the aggregate, and were based on a biological study of cockroach aggregation. The model was used to predict the probability of a swarm aggregating, without capturing spatial information. Despite the limitation of assuming that an aggregate can only grow or shrink linearly, they conclude that robots need a minimal speed or sensor range to enable aggregation.

Rouff et al. compared four different specification formalisms (\gls{csp}, \gls{wsccs}, Unity Logic, and X-Machines) for specifying and verifying emergent swarm behaviour~\cite{rouff2004formal}. They concluded that, a blending of these formalisms offered the best approach to specify emergent swarm behaviour as none was sufficient in isolation. It is thus clear that only the use of multiple, specialised tools and methodologies can achieve a high level of confidence in software for robot swarms, and more generally, software for robotic systems~\cite{hoare2009viewpoint}. For example, the NASA Remote Agent uses specialized languages for each of the planner, executive and fault diagnosis components~\cite{simmons_towards_2000}. We explore the use of integrated formal methods for the specification and verification of robotic systems in \S\ref{sec:ifm}.

\subsubsection{Teams}
\label{sec:teams}

A robot team is similar to a swarm, but contains robots with different capabilities or behaviours to one another. For example, a search and rescue robot team might contain an autonomous aircraft to search the area, a wheeled robot to carry equipment, and a robot with caterpillar tracks that is capable of moving obstacles. The heterogeneity of the robot's capabilities and behaviour increases the likelihood that they will each need different formal methods during verification. Linking verification at the micro-level to the macro-level is, therefore, challenging. We discuss the use of integrated formal methods in more detail in \S\ref{sec:ifm}.

The work in~\cite{kloetzer2011multi} presents a methodology for automating the development of robot teams, using \gls{ltl-x}. They model-check a transition system describing the behaviour of the robot team for execution traces that satisfy an \gls{ltl-x} formula. This trace is mapped to the communication and control strategy for each robot. \gls{ba} then produce a communication strategy that reduces the number of `stop-and-wait' synchronisations that the team has to perform. 

Robot teams often include robots with different capabilities, which produce different patterns of interaction. The work in~\cite{hilaire2007formal} captures part of the Satisfaction-Altruism Model~\cite{Simonin2000} to provide a formal model of the different and changing roles that different agents in a team can play. Each role defines a pattern of interaction. They propose using their model for specification and verification of a robotic team, and eventual refinement to code. Their model is formalised using an integrated formal method (discussed in \S\ref{sec:ifm}) that combines Object-Z and Statecharts~\cite{Harel1987}. 

Previous work in this area has assumed that a robot and a human are in close contact and that the robot can continually observe the human's behaviour, allowing it to predict the human's plan. Talamadupula et al.~\cite{talamadupula2014coordination} describe a \gls{fol} formalisation of beliefs and intentions that allows a robot to predict another agent's plan. They explore a human-robot cooperation task where the robot must predict a human's plan based on mutually understood beliefs and intentions. The \gls{fol} predicates representing the beliefs and intentions of another agent (human or robotic) are input to the robot's planning system, which predicts the other agent's plan. An extension caters to incomplete knowledge of the other agent's goals.
  
\subsection{Self-Adaptive and Reconfigurable Systems}
\label{sec:reconfig}

Ensuring fault tolerance of robotic systems that operate in hazardous environments, potentially isolated from human interaction, is crucial. These systems must be able to diagnose and to reconfigure themselves in order to adapt to changes in their requirements or operating environment. Reconfiguration can help maintain quality of service or meet functional objectives~\cite{Weyns2010}. This is particularly relevant for robotic systems that are deployed in hostile environments, such as in space or the deep ocean~\cite{Gao2016,tarasyuk2012formal}. \textit{Rational} autonomy, where a system can explain its reasoning, is crucial for safe reconfiguration. Therefore, modelling the motivations and decisions of the system is an important line of current work. Ideally, this should be done in a way that enables formal verification of the autonomous decisions~\cite{fisher_verifying_2013}. This is discussed in more detail in \S\ref{sec:agents}.

A self-adaptive system continually alters its behaviour in an environmentally-driven feedback loop. The findings of a literature survey indicated that while there are no standard tools for the formal modelling and verification of self-adaptive systems; 40\% of the studies they surveyed used formal modelling or validation tools, and 30\% of those tools were model-checkers~\cite{Weyns2012}. 
Architecture-based self-adaptive systems are a specific class of self-adaptive system where the system reasons about a model of itself and the environment, and adapts according to some adaptation goals. In this case, the feedback loop consists of Monitor, Analyse, Plan and Execute components and runtime models which provide Knowledge (MAPE-K) \cite{iglesia2015mape}. A series of MAPE-K templates has been formally verified using \gls{ta} and \gls{tctl} to aid in development.

Related to the notion of adaptation is the concept of a reconfigurable system, which senses the environment and makes a \textit{decision} about how best to reconfigure itself to suit changes in its requirements or the physical environment. Reconfiguration is essential for ensuring the fault-tolerance of a safety-critical robotic system~\cite{tarasyuk2012formal}. 
The reconfigurable systems literature focuses on two research questions: (1) how to specify and analyse a reconfigurable system~\cite{Morse2016}, and (2) how to compare different configurations of the system~\cite{Weyns2010, Morse2016}? Therefore, it seems crucial that formal techniques applied to reconfigurable systems are able to tackle these questions.

The development of reconfigurable machines is an area where a substantial amount of research has been carried out with respect to the hardware for such machines, but developing software that can autonomously reconfigure the system is a challenge for the entire community \cite{bi2008development}. The development of such systems places a heavy emphasis on designing these systems to be modular, thus making reconfiguration a more approachable endeavour. 

Reconfigurability can be acheieved by developing a flexible control system that can reconfigure itself when a fault is detected~\cite{braman2007safety}. Specifying this can be challenging, however. The work in~\cite{iftikhar2014activforms} presents a collection of \gls{ta} models for different modes of operation, and a module that enables them to be swapped during their execution on a \gls{vm}. Executing the \gls{ta} models ensures that the model's verified behaviour is the program's runtime behaviour. 
Recent work includes a methodology for verifying reconfigurable timed net condition/event systems \cite{hafidi2018methodology} and stochastic petri nets to verify reconfigurable manufacturing systems \cite{tigane2016reconfigurable}.

One avenue of research suggests providing (both semi-formal and formal) models to be used at runtime~\cite{Cheng2014}. This agenda considers approaches such as automatic test case generation and formal model-checking. This relies on the fact that many variables that are unknown during the design of a system can be quantified at runtime, which helps to control the problems of state space explosion. This has the benefit of providing online assurance of an adaptive system, \textit{while} it is adapting; which can improve the trust in results regarding safety.

Decentralised systems can bring their own set of challenges for adaptation and reconfigurability. Iftikhar and Weyns~\cite{iftikhar2012case} use a traffic monitoring system as their case study, where a collection of cameras monitor the traffic on a stretch of road. When a traffic jam is detected, the cameras collaborate in a master-slave architecture to report information about the traffic. This self-adaptation is managed without a central control unit, to avoid the communications bottleneck this would cause. They model this system using \gls{ta} and describe invariants and correctness properties using \gls{tctl}. Related work utilised a Monte Carlo approach to probabilistically verify a system of cooperating agents (autonomous vehicles) that implement a policy-based collision avoidance algorithm~\cite{pallottino2007decentralized}.

\subsection{Summary of Internal Challenges}
\label{sec:challengesSummary}

This section identified three internal challenges to the specification and verification of robotic systems related to how the robotic system is designed and built. These challenges are: agent-based, multi-robot, and reconfigurable systems. 

Agent-based systems are not the only model of autonomy, but they encapsulate the description of autonomous behaviour into one component, and can be used to model interactions with other actors and the environment. A \textit{rational} agent can explain its reasoning, which can be helpful during verification and in tackling the challenge of providing sufficient evidence for a certification body or to gain public trust (recall \S\ref{sec:trustAndCertEvidence}). Ensuring that agents are verifiable is a focus in the literature.

Multi-robot systems -- both homogeneous swarms and heterogeneous teams -- provide an interesting challenge to robotic systems. They provide resilient architectures; but their decentralisation brings specific challenges for formal verification.
Reconfigurable systems are key to dealing with changing environments and mission goals. However, ensuring that they reconfigure themselves in a safe way is a challenge, and the key focus in the literature. 

These are the internal challenges that we derived from the literature, but of course, there may be others. For example, machine learning systems are becoming more prevalent in autonomous robotic systems and are notoriously difficult to formally verify, although recent work has emerged~\cite{selsam2017developing, HuangKWW17}.

\section{Formalisms for Robotic Systems}
\label{sec:formalisms}

This section discusses the formalisms found in the literature, following the methodology from \S\ref{sec:scope}, that have been used to specify or verify robotic systems. Table~\ref{tab:formalismSummary} summarises the categories of formalisms being used to specify both the system and the properties being checked\footnote{We note that these are not definitive categories, but represent common terms for each type of formalism.}. We explicitly display the formalisms used to specify both the system and the properties because it illustrates the (well known) links between types of formalism (e.g. state-transition systems and temporal logics); it also highlights potential gaps in the literature or areas to focus on when producing tools. 

We found that systems were normally specified as state-transition systems and the properties using a logic -- typically, a temporal logic. The `Other' category contains those which do not fit in the other categories but are not numerous enough to warrant their own. Examples include the \textsc{relax}~\cite{whittle2009relax}requirements language and the Quartz language used with the Averest framework~\cite{proetzsch2007formal}.

\begin{table}
\centering
\scalebox{0.8}{
\begin{tabular}{|c l ||m{5.5cm}|c||m{5.5cm}|c| }
\hline
 &  & \multicolumn{2}{c||}{\textbf{System}} & \multicolumn{2}{c|}{\textbf{Property}} \\
\hline
\multicolumn{2}{|c||}{\textbf{Formalism}} & \textbf{References} & \textbf{Total} & \textbf{References} & \textbf{Total}  \\ \hline \hline

\multicolumn{2}{|l||}{Set-Based} & \cite{gudemann2008specification}, \cite{hilaire2007formal}, \cite{liang2009software}, \cite{Weyns2010}, \cite{weyns2012forms} & 5 &  & 0 \\ \hline \hline

\multicolumn{2}{|l||}{State-Transition} &\cite{abdellatif2012rigorous}, \cite{basu2008incremental}, \cite{basu2011rigorous}, \cite{bensalem2009designing}, \cite{brambilla2012property}, \cite{celaya2007modeling}, \cite{chen2012formal}, \cite{chen2013formal}, \cite{correll2007modeling}, \cite{costelha2007modelling}, \cite{dixon2011towards}, \cite{dixon2012towards}, \cite{falcone2011runtime}, \cite{filippidis2012decentralized}, \cite{guo2013revising}, \cite{hess2014formal}, \cite{iftikhar2012case}, \cite{iftikhar2014activforms}, \cite{iglesia2015mape}, \cite{Izzo2016}, \cite{karaman2009sampling}, \cite{kloetzer2011multi}, \cite{Konur2010}, \cite{konur2012analysing}, \cite{kress2007s}, \cite{kress2011correct}, \cite{molnar2009system}, \cite{raman2011analyzing}, \cite{risler2008formal}, \cite{webster_formal_2014}, \cite{webster_toward_2016}, \cite{winfield2008modelling}, \cite{ziparo2008petri} & 33 & & 0 \\ \hline \hline

\multicolumn{2}{|l||}{Logics} &  & 6 & & 32 \\ \hline 

	&	Temporal Logic &  & 0 & \cite{bonfe2012towards}, \cite{brambilla2012property}, \cite{Dennis2015}, \cite{Dennis2016a}, \cite{dixon2011towards}, \cite{dixon2012towards}, \cite{filippidis2012decentralized}, \cite{fisher_verifying_2013}, \cite{guo2013revising}, \cite{huang2014rosrv}, \cite{iftikhar2012case}, \cite{iftikhar2014activforms}, \cite{iglesia2015mape}, \cite{Izzo2016}, \cite{karaman2009sampling}, \cite{kloetzer2011multi}, \cite{Konur2010}, \cite{konur2012analysing}, \cite{kress2007s}, \cite{kress2011correct}, \cite{massink2013use}, \cite{molnar2009system}, \cite{proetzsch2007formal}, \cite{raman2011analyzing}, \cite{webster2011formal}, \cite{webster_formal_2014}, \cite{webster_toward_2016} & 27 \\ \hline	

	&	Dynamic Logic & \cite{kouskoulas2013certifying}, \cite{mitsch2013provably} & 2 & \cite{kouskoulas2013certifying}, \cite{mitsch2013provably} & 2 \\ \hline
	
	&	Other Logics & \cite{bonfe2012towards}, \cite{dylla2008approaching}, \cite{gjondrekaj2012towards}, \cite{talamadupula2014coordination} & 4 & \cite{dylla2008approaching}, \cite{gjondrekaj2012towards}, \cite{talamadupula2014coordination} & 3 \\ \hline \hline
	
\multicolumn{2}{|l||}{Process Algebra} & \cite{akhlaki2007methodological}, \cite{massink2013use}, \cite{obrien_automatic_2014} & 3 & \cite{obrien_automatic_2014} & 1 \\ \hline \hline

\multicolumn{2}{|l||}{Ontology} & \cite{alsafi2010ontology}, \cite{molnar2009system}, \cite{PRESTES20131193}, \cite{tenorth2009knowrob} & 4 & & 0 \\ \hline \hline

\multicolumn{2}{|l||}{Other} & \cite{Dennis2015}, \cite{Dennis2016a}, \cite{fisher_verifying_2013}, \cite{proetzsch2007formal},  \cite{webster2011formal} & 5 & \cite{abdellatif2012rigorous}, \cite{basu2008incremental}, \cite{basu2011rigorous}, \cite{bensalem2009designing}, \cite{chen2012formal}, \cite{chen2013formal}, \cite{falcone2011runtime}, \cite{whittle2009relax} & 8 \\ \hline

\end{tabular}}
\caption{\label{tab:formalismSummary} Summary of the types of formalisms found in the literature for specifying the system and the properties to be checked. Note: some references appear in one half of the table but not the other because they only describe the specification of the system (or vice versa). Logics have been further subdivided into Temporal, Dynamic, and Other Logics.}
\vspace{-1.5em}
\end{table}

Table~\ref{tab:toolSummary} summarises the wide variety of formal analysis tools that were found using our methodology. Several studies use tools that specify properties in temporal logic; e.g. \Uppaal{}~\cite{aniculaesei2016towards}, PRISM~\cite{Brambilla2014}, etc. The number of tools in use for temporal logic is commensurate with their prevalence as illustrated in Table~\ref{tab:formalismSummary}. 

\begin{table}
\centering
\scalebox{0.8}{

\begin{tabular}{|l|l|l|l|l|}
\hline
\textbf{Type of Tool}           & \textbf{Tool}  & \textbf{References} & \textbf{Total} & \textbf{Type Total}    \\ \hline \hline

\multirow{9}{*}{Model-Checkers} & \Prism{} &	\cite{Konur2010}, \cite{massink2013use}, \cite{konur2012analysing}, \cite{brambilla2012property} &	4  &  \multirow{9}{*}{25} \\ \cline{2-4}
								& \NuSMV{} &	\cite{dixon2011towards}, \cite{dixon2012towards}	&	2 &\\ \cline{2-4}
								& \Uppaal{} &	\cite{iftikhar2012case}, \cite{iftikhar2014activforms},	\cite{iglesia2015mape} & 3 &\\ \cline{2-4}
								& SAL	& \cite{hilaire2007formal}	&	1 & \\ \cline{2-4}
								& SPIN &	\cite{webster_formal_2014}, \cite{webster_toward_2016},  \cite{dantam2012robust}, \cite{filippidis2012decentralized}, \cite{webster2011formal} & 5 &\\ \cline{2-4}
								& Beryl &	\cite{proetzsch2007formal}, \cite{proetzsch2007behaviour}	& 2	 &\\ \cline{2-4}
								& Aldebaran	 &	\cite{chkouri2008translating}	 &		1 &\\ \cline{2-4}
								& Dfinder &		\cite{bensalem2011formal}, \cite{basu2011rigorous}, \cite{bensalem2009designing}, \cite{abdellatif2012rigorous}	 &		4& \\ \cline{2-4}	
								& Unspecified & \cite{chen2012formal}, \cite{bordini2009formal}, \cite{karaman2009sampling} & 3  &\\ \hline \hline

\multirow{2}{*}{Program Model Checkers} & AJPF  		& \cite{fisher_verifying_2013}, \cite{Dennis2016a}, \cite{Dennis2015}, \cite{webster2011formal}          & 4     & \multirow{2}{*}{7} \\ \cline{2-4}
                                        & MCMAS 		& \cite{Kouvaros2017}, \cite{choi2015verification}, \cite{molnar2009system}          & 3     &                    \\ \hline \hline

\multirow{2}{*}{Theorem Provers}      & KeyMaera &		\cite{mitsch2013provably}, \cite{kouskoulas2013certifying}	 &		2  &  \multirow{2}{*}{3} \\ \cline{2-4}
& SteP	 & \cite{hilaire2007formal}	&	1 & \\ \hline \hline

\multirow{10}{*}{Others} & Bio-PEPA Tool Suite &		\cite{massink2013use}	 &		1  & \multirow{10}{*}{14}\\ \cline{2-4}
& TmeNET &		\cite{costelha2007modelling}	 &		1 & \\ \cline{2-4}
& TuLiP	 &	\cite{kress2011correct}	 &		1 & \\ \cline{2-4}
& LTLMoP &		\cite{kress2011correct}, \cite{raman2011analyzing}	 &		2 & \\ \cline{2-4}
& Alloy & \cite{bonfe2012towards}, \cite{gudemann2008specification} & 2 & \\\cline{2-4}
& Evaluator & \cite{basu2008incremental} & 1 & \\ \cline{2-4}
& minisat & \cite{basu2008incremental} & 1 & \\\cline{2-4} 	
& MissionLab (VIPARS) & \cite{obrien_automatic_2014} & 1 & \\\cline{2-4} 
& RV-BIP & \cite{falcone2011runtime} & 1 & \\\cline{2-4}
& Community Z Tools & \cite{liang2009software},\cite{Weyns2010}, \cite{weyns2012forms}& 3 & \\

\hline

\end{tabular}
}

\caption{\label{tab:toolSummary}Summary of the formal verification tools, and the type of tools, identified in the core set of 63 papers surveyed. Some approaches use multiple tools, often to verify different parts of a system with the most suitable tool chosen for each individual component of the system.}
\vspace{-1.5em}
\end{table}

In this section, we describe the formalisms used to specify and verify robotic systems. \S\ref{sec:modelSpec} discusses set-based formalisms. \S\ref{sec:automota} describes approaches using automata. \S\ref{sec:logics} describes approaches using logics (temporal, dynamic and others). \S\ref{sec:procesAlgebra} discusses process algebras. \S\ref{sec:ontologies} outlines ontologies that have been used, and \S\ref{sec:others} describes other existing formalisms. Finally, \S\ref{sec:formalismsSummary} summarises.

\subsection{Set-Based Formalisms}
\label{sec:modelSpec}

Set-Based formalisms (such as the B-Method and Z) specify a system using set-theoretic representation and manipulation of data. They are well suited to capturing complicated data structures, but usually only provide limited, if any, features for capturing behaviour. 

As mentioned in \S\ref{sec:reconfig}, self-adaptive (or reconfigurable) behaviour is often key for autonomous robotic systems. Weyns et al. describe a formal reference model for self-adaptive systems, called FORMS~\cite{Weyns2010}, which provides a Z model that describes an arbitrary self-adaptive system. They suggest that this model could be used in conjunction with various existing tools to type check the adaptations to ensure that they are valid, visualise the model using animators, automatically generate test cases, or to transform the model into other notations for other types of analysis. The model provides a method of describing a self-adaptive system to enable communication about the adapting architecture during system design and comparison of different architectural configurations. 

Liang et al.~\cite{liang2009software} use the Jaza animator for Z specifications, combined with a Java debugger, to provide a runtime monitoring system. A Z specification describes the system's behaviour, which is animated and compared to the running program via the \textbf{\textit{jdb}} Java debugger to check that the running program implements the Z specification. They demonstrate this approach on a robotic assembly system, designed to build robots for the NASA ANTS project\footnote{\url{https://attic.gsfc.nasa.gov/ants/} }. Two advantages of combining the Z animator with a Java debugger are that it keeps the monitor and program code separate, and that the approach doesn't require any alterations to the monitored program. 

In~\cite{tarasyuk2012formal}, Event-B specifications are combined with probabilistic properties to derive reconfigurable architectures for an on-board satellite system. Event-B evolved from the B-Method and is more suitable to modelling the dynamic behaviour of a system than its predecessor \cite{abrial_modeling_2010}. The combination of these formalisms allows the models of reconfigurations to be checked using PRISM for both the derivation (via refinement) of the system from its specification, and the probabilistic assessment of their reliability and performance. 

\subsection{State-Transition Formalisms}
\label{sec:automota}

Petri Nets, \glspl{fsm} and \gls{fsa} are approaches to specifying behaviour as a state-transition system. Formalisms for discrete-event systems include those capturing time (e.g. \gls{ta} or Timed Petri Nets) and probabilistic transitions (e.g. \gls{pfsm} or \gls{gspn}). State-transition systems specify behaviour during the design phase, which is checked for properties like deadlock freedom; or used as an input to a tool that usually checks them for properties described in another formal language (e.g. a temporal logic, see \S\ref{sec:tempLogic}). There is a close link between varieties of temporal logic and varieties of \gls{fsa}~\cite{Vardi95}, and many works combine both.

Petri Nets have had some use in modelling robotic systems. The work in~\cite{celaya2007modeling} uses them to capture the abstract architecture of multiple reactive agents and they are analysed for deadlock freedom. In~\cite{ziparo2008petri}, Ziparo et al. extend Petri Nets to capture robot plans. This extension to Petri Nets is a set, $G$, of Goal Markings, which is proper subset of the reachable markings for the Petri Net that represents the desired result of the plan. The Petri Net Plans can be executed to find a sequence of transitions from the initial to the goal markings.

Costelha and Lima~\cite{costelha2007modelling} use four layers of Petri Nets to model a robot's task plans, each layer describes a distinct element at a different level of abstraction. They use \glspl{mopn} to specify the robot's environment to enable a more realistic model. Then they capture the actions that a robot can take, the task plans, and the roles of various robots working together, using \glspl{gspn}. These models can be analysed for deadlocks and resource usage.

The \Uppaal{} model-checker uses networks of \gls{ta} to describe input models, which are checked against temporal logic properties.
The work in~\cite{Halder2017} targets the ROS middleware framework, capturing the communication between nodes using \gls{ta}. The low-level information (e.g. queue sizes and time outs) allows the verification of safety and liveness properties with \Uppaal{}. In~\cite{iftikhar2012case}, \gls{ta} are used to model a decentralised traffic monitoring system, and \gls{tctl} to describe system invariants and correctness properties of the system's self-adaptive behaviour which are verified using \Uppaal{}.

The work in~\cite{Lyons2016} uses \gls{fsa} and a Gaussian model of the environment, based on a probability distribution of obstacle positions. These models are combined in MissionLab, a graphical robot mission designer. The robot \gls{fsa} models are automatically translated into MissionLab's internal process algebra, \gls{pars} which will be discussed  in \S\ref{sec:procesAlgebra}. 

As described in \S\ref{sec:agents}, Hoffmann et al.~\cite{Hoffmann2016} extend \glspl{pfsm} with a weight function to map weights onto actions -- which they call an \gls{apfsm}. Because of the addition of the weight function, an \gls{apfsm} is a discrete-time Markov chain. They use this formalism to describe autonomous behaviour and \Prism{} to verify properties about a pilotless aircraft on a foraging mission, with probabilistic modelling used to capture the physical environment. 
Markov chains have also been used to build models for predicting the likelihood of a robot swarm aggregating~\cite{correll2007modeling}. Each robot is represented by a Markov chain, where the probabilities to join or leave an aggregate are a function of the size of the aggregate (larger aggregates are preferred). % already said this: The probabilities are based on the results of a biological study of cockroach aggregation. 
The authors of~\cite{correll2007modeling} find that their model correlates closely with results from simulations. 

\subsection{Logics}
\label{sec:logics}
Logics are the second most prevalent formalism found during our literature search, the majority of which are some variety of temporal logic. We also found work using dynamic logics and a variety of logics that didn't warrant their own category (e.g. FOL). Temporal logics are discussed in \S\ref{sec:tempLogic}, dynamic logics in \S\ref{sec:dynamicLogic}, and other logics in \ref{sec:otherLogics}.

\subsubsection{Temporal Logics}
\label{sec:tempLogic}
~\\ \noindent
Temporal logics are used for specifying dynamic properties about a system over linear or branching time, which feature heavily in the literature (Table~\ref{tab:formalismSummary}). There are a variety of temporal logics: \gls{ltl} and \gls{ctl} deal with events occurring next, globally, or eventually; extensions like \gls{ptl} and \gls{pctl} add notions of probability. Temporal logics have been used extensively to address the challenges that we identified in \S\ref{sec:extChallenges} and \S\ref{sec:intChallenges}.

Modelling a robot's environment is a key challenge when building robotic systems (\S\ref{sec:environment}). One approach~\cite{guo2013revising} uses a \gls{ba} to model the environment and synthesises a motion controller by model-checking it to find an accepting path that satisfies an \gls{ltl} task specification. The \gls{ba} is updated at runtime with new environmental information and checked again to revise the motion controller. 

The work in~\cite{Izzo2016} presents a probabilistic approach to modelling a \gls{bdi} robotic agent and its environment. The agent can be captured as either a \gls{dtmc} or a \gls{mdp}. Then, they use \gls{pctl} to specify properties, which are model-checked, using PRISM. They apply this technique to an autonomous mine detector, showing how their approach can be used for both design-time checking and runtime monitoring (which we discuss in \S\ref{sec:rvmonitor}). The runtime safety monitors for autonomous systems presented in~\cite{Machin2015} are specified in \gls{ctl}. The work in~\cite{Desai2017}, aimed generally at safe robotics, captures assumptions about the real world using \gls{stl} and uses them for runtime monitoring. Another example, this time aimed at robot swarms~\cite{Lopes2016}, models the system's potential behaviour and constrains it according to \gls{sct} specifications of safe behaviour.

%Safety, aircaft, ltl
Several approaches use temporal logic to provide certification evidence or build public trust in autonomous robotic systems. The work in~\cite{webster2011formal, Webster2014} captures air safety rules and assumptions using \gls{ltl} (extended to capture \gls{bdi} agent beliefs) which are used to verify that an autonomous pilotless aircraft follows the rules of the air in the same way as a pilot. Similarly, \cite{Dixon2014} and \cite{Gainer2017} present an approach for automatically building \gls{ptl} models of the safety rules and environment of a robotic domestic assistant. These models are checked against a probabilistic description of the robot's environment (a sensor-equipped house in the UK) to ensure that the rules are sufficient to keep a human safe. An extension~\cite{webster_toward_2016} checks models of the rules against environment models based on data collected from a person living in the house for several days. These verification results improve the confidence in the safety of the robot's high-level decision making.

Enabling robots to deal with ethical choices is a current open challenge. One approach presents a \gls{bdi} language, \Ethan{}~\cite{Dennis2016a}, for reasoning about ethical principles. It can be verified, using \gls{ajpf}, that when an \Ethan{} agent chooses a plan, all other such plans are ethically worse.

Another popular approach is synthesising behavioural models that satisfy temporal logic properties. For example, in~\cite{karaman2009sampling} motion plans are built incrementally that guide the robot to its goal position, while satisfying properties that are specified in a deterministic subset of \Mucalc{}. There is potential for using this approach at runtime; a plan can be built from a system of over 1000 states in $\sim$3.5 seconds. In~\cite{kress2011correct}, \gls{ltl} specifications are used to synthesise robot motion automata. This approach mediates state explosion by planning a short distance ahead, repeating after each plan is complete. The work in~\cite{raman2011analyzing} presents a tool that generates hybrid robot controllers from \gls{ltl} specifications, which can be used in simulation or with real robots.

Temporal logics have also been used to verify particular robotic software frameworks. The existing tool-chain for \Genom{} has been extended to generate Fiacre models of a system, which can then be model-checked against \gls{ltl} properties~\cite{Foughali2016}. Another approach models ROS nodes, and the communication between them, as \gls{ta}; which are
checkd against \gls{tctl} properties in \Uppaal{}~\cite{Halder2017}. They verify the lack of message buffer overflows for a selection of queue sizes in the open-source robotic application Kobuki.

Temporal logics have been used to help overcome some of the challenges in developing swarm robotic systems. Winfield et al. describe a \textit{dependable swarm}~\cite{winfield_formal_2005}: a distributed multi-robot system, based on the principles of swarm intelligence, that we can rely on the behaviour of. They advocate the use of temporal logic for specifying both safety and liveness properties of a simple wireless connected swarm. As part of their approach they discretise the robot's environment to a grid in order to simplify its specification. This is a common approach taken when using formal methods, since modelling continuous elements of the system is difficult. In a companion paper~\cite{dixon2012towards} they apply model-checking techniques to the alpha swarm navigation algorithm.

As mentioned in \S\ref{sec:swarms}, swarm robotic systems exhibit both macroscopic and microscopic properties. The authors of~\cite{Moarref2017} propose a novel specification language to allow the explicit specification both of the behaviour of the whole swarm and of individual robots. Their language captures the swarm's environment as a \textit{region graph}, describes the swarm's movement using \textit{region propositions}, and specifies the swarm's macroscopic and microscopic behaviour (separately) using \gls{ltl}. These components are combined to produce decentralised controllers for the robot swarm. 

In contrast, \cite{filippidis2012decentralized} presents a technique for dealing with decentralised control in multi-agent systems that keeps the specifications local to each robotic agent. Each local specification is written in \gls{ltl}, with their mutual satisfiability not guaranteed beforehand. The local specifications are exchanged by the robots at physical meeting points and model-checked for mutual satisfiability. However, this technique assumes that dependencies between robotic agents are sparse, so it may not scale well in robot swarms or teams with a larger number of dependencies.

Probabilistic temporal logics have proved useful in modelling robot swarms~\cite{Liu2007}. The work in~\cite{Konur2010} models the probabilistic state machine from~\cite{Liu2007}, which is checked in \Prism{} for satisfaction of \gls{pctl} properties. Since each robot in the swarm is characterised by the same state machine, a \emph{counting abstraction} is used to reduce the state space; but, this abstraction only works for homogeneous swarms. 

The work in~\cite{Brambilla2014} presents a top-down swarm development methodology that begins with a \gls{pctl} specification of the swarm's macroscopic requirements. Successively, more concrete models of the swarm are developed, including simulation and field test implementations. Each step is checked against the previous steps to ensure that the swarm's behaviour matches the initial specification. However, this approach only focusses on the macroscopic level of the behaviour of the swarm, other approaches consider the microscopic level as well. 

\subsubsection{Dynamic Logic} 
\label{sec:dynamicLogic}
~\\ \noindent
Dynamic logic is an extension of modal logic that can be used to specify and reason about properties of dynamic systems thus they are useful for specifying robotic systems.

Mitsch et al. use \gls{dl}~\cite{Platzer2007}, which was designed for the specification and verification of hybrid programs, to describe the discrete and continuous navigation behaviour of a ground robot \cite{mitsch2013provably, Mitsch2017}.  Mitsch et al. use the hybrid theorem prover KeYmaera~\cite{mitsch2013provably} (and its sucessor, KeYmaera X~\cite{Mitsch2017}) to prove safety (and liveness) properties about a robot's behaviour, written in \gls{dl}.  They also describe an automatic synthesis, from the verified \gls{dl} model, to runtime monitors for safety properties \cite{Mitsch2017}. 

Other work uses quantified differential dynamic logic Q\gls{dl}  to analyse a control algorithm of a surgical robot \cite{kouskoulas2013certifying}. They also used a variation of KeYmaera, called KeYmaeraD, for proof support.

\subsubsection{Other Logics} 
\label{sec:otherLogics}
~\\ \noindent
We identified a number of other logics that were not numerous enough to for their own category. 

Dylla et al.~\cite{dylla2008approaching} formalise the basic concepts of football theory (from a leading football theory manual) using Readylog -- which is a variant of the logic programming language, Golog. They use this formalisation to formally describe football tactics, which they implement on football-playing robots for the RoboCup competition. 
  
Gjondrekaj et al.~\cite{gjondrekaj2012towards} use a formal language that has been designed to capture properties about distributed systems, \Klaim{}, its stochastic extension, \StoKlaim{}, and their related tool set. Their application domain is distributed robotic systems and they describe the modelling of three homogeneous robots, under distributed control, cooperating to transport objects.

Talamadupula et al.~\cite{talamadupula2014coordination} use \gls{fol} to capture an agent's beliefs and intentions. They use this to enable the prediction of other agents' plans.

\subsection{Process Algebras}
\label{sec:procesAlgebra}

Process algebraic approaches define the behaviours of a system in terms of events and the interactions of processes. They are well suited to specifying concurrent systems. Process algebras can capture (discrete) time, but they seldom capture probabilistic behaviour. 

Multi-agent systems can be described by a combination of the process algebra \gls{fsp} and \gls{piadl}~\cite{Akhtar2014}. \gls{fsp} specifies the required safety and liveness properties, which are transformed into \glspl{lts}; then the agent program and architecture (writted in \gls{piadl}) are checked to see if they satisfy these properties.

The Bio-PEPA process algebra has been used to model the microscopic behaviour of a robot swarm \cite{massink2013use}. It enables several different analysis methods using the same specification. They describe a specification of a robot swarm performing a foraging task where a team of three robots is required to collect each object, and the teams vote on taking either a long or short path between the start and pick-up points. They present results about the model using stochastic simulation, statistical model-checking, and ordinary differential equations. This collection of results is compared with an existing analysis of the same case study; the comparison shows that Massink et al.'s approach provides similar results, but enables a wider range of analysis methods from a single specification.

RoboChart provides a formal semantics, based on \gls{csp}~\cite{Hoare1978}, for a timed state machine notation~\cite{Ribeiro2017}. This is a similar approach to the (non-formal) state chart notations ArmarX~\cite{Wachter2016} and rFSM~\cite{Merz2006}, described in \S\ref{sec:nonFormal}.
RoboChart is supported by an Eclipse-based environment, RoboTool~\cite{Miyazawa2017}, which allows the graphical construction of RoboChart diagrams that it can be, automatically, translated into CSP, in the form described above. In \gls{csp}, events are instantaneous. RoboChart allows the specification of explicit time budgets and deadlines, which RoboTool translates into Timed~\gls{csp}~\cite{Reed1986}.  

The \gls{csp} model-checker, \gls{fdr}~\cite{GibsonRobinson2014}, can be opened from within RoboTool to enable quick checking of the timed and untimed properties of the RoboChart model. RoboTool automatically generates basic assertions (to check for deadlock, livelock, timelock, and non-determinism) for \gls{fdr}. These can be edited, but this requires some skill with \gls{csp} and knowledge of the model, which many software engineers do not have. The ability to graphically edit and visualise the state machines is a key aspect of being able to integrate this notation into an existing engineering process. 

The timed process algebra, CSP+T (which extends \gls{csp} with a notion of time), has also been used to capture the real-time aspects of robotic systems. The work in~\cite{akhlaki2007methodological} derives correct and complete CSP+T specifications from a description of a system in the \gls{uml} profile UML-RT -- designed for developing real-time systems. This work maps each subsystem in the UML-RT description of the system to a CSP+T process, and composes them in parallel. They demonstrate this approach on a two-armed industrial robot. However this mapping appears to be a manual, informal process. 

As previously mentioned in \S\ref{sec:automota}, the graphical robot mission design tool MissionLab contains a process algebra, \gls{pars}. The work in~\cite{Lyons2016} describes a robot's behaviour using \gls{fsa} and its environment using a Gaussian model, but MissionLab automatically translates these into \gls{pars} for checking the  robot's behaviour within the given environment.

\subsection{Ontologies}
\label{sec:ontologies}

Ontologies act as a body of knowledge and provide a vocabulary for a given domain~\cite{haidegger2013applied}. They formally specify the ``key concepts, properties, relationships and axioms of a given domain'' and enable reasoning over this knowledge to infer new information~\cite{PRESTES20131193}.

For autonomous robotics, ontologies have been used to describe the robot environment, describe and reason about actions, and for the reuse of domain knowledge~\cite{PRESTES20131193}. Ontologies also have the benefit of being able to capture elements of Human-Robot Interaction. Despite finding ontologies for autonomous robotics systems, our literature search found little evidence of available tool support. 

The IEEE-RAS working group \textit{Ontologies for Robotics and Automation} has published an ontology\footnote{IEEE 1872-2015 - IEEE Standard Ontologies for Robotics and Automation \url{https://standards.ieee.org/standard/1872-2015.html}} to explicitly and formally specify the shared concepts within robotics and automation~\cite{PRESTES20131193, haidegger2013applied}. 

\textsc{KnowRob} is a knowledge processing system for autonomous personal robotic assistants~\cite{tenorth2009knowrob} where knowledge is represented in description logic using \gls{owl}. Their system can be used with ROS and integrates encyclopaedic knowledge, an environment model, action-based reasoning and human observations. 
In \cite{alsafi2010ontology}, they use ontological knowledge, represented using the \gls{owl}, of a manufacturing environment to enable reconfiguration without human intervention. This work uses \gls{owl} reasoners in order to decide how best to reconfigure the system.

\subsection{Other Formalisms}
\label{sec:others}

Besides those described in the previous subsections, a number of other formalisms and approaches feature in the literature. Some of these formalisms are used to specify the system and others are used to specify the properties the system is being checked for. In the literature we found, each of the studies that used `other' formalisms only specify either the system or the properties, not both.

As previously mentioned in \S~\ref{sec:agents}, autonomous robotic systems are often programmed using agents. Several works we found~\cite{Dennis2015,Dennis2016a,fisher_verifying_2013,webster2011formal} make use of the \gls{ajpf}~\cite{Dennis2012} program model checker. The properties being checked by \gls{ajpf} are described in temporal logic, but the system is a \gls{bdi} agent program written in \Gwen{}. The \gls{ajpf} translates agent programs into a formally-defined internal representation to enable model checking.

The work in~\cite{proetzsch2007formal} describes the verification of safety properties using the Averest framework. Robotic systems are described using the Quartz language and the safety properties are specified in $\mu$-calculus. Quartz\footnote{\url{http://www.averest.org/\#about_thesynchronouslanguagequartz}} is a synchronous language that allows the developer to describe, for example, parallel execution and nondeterministic choice.

Several papers use the \gls{bip} framework~\cite{Basu2006}, which uses familiar \glspl{fsm} to specify a system but provides the D-Finder tool~\cite{Bensalem2011DFinder} to check for deadlocks and other safety properties. Some work shows how \gls{bip} can be used to specify and verify robotic software~\cite{basu2008incremental, basu2011rigorous, bensalem2009designing}. In addition, \Rvbip{}~\cite{falcone2011runtime} is a tool that produces a runtime monitor as an ``additional component in a \gls{bip} system'' that can check an \gls{lts}-based specification at runtime. Finally, the work in \cite{abdellatif2012rigorous} provides a synthesis technique to generate robotic software from \gls{bip} models.

Another synthesis technique is presented in \cite{chen2012formal} and \cite{chen2013formal}, which tackle robotic teams' communication strategies. The properties that the robotic team must exhibit (that is, the tasks) are specified using a (formally defined) regular expression language. These specifications are used to synthesise robot controllers, as \gls{fsa}.

\textsc{relax} is a requirements language for dynamic self-adaptive systems that enables developers to indicate requirements that may be relaxed at run-time~\cite{whittle2009relax}. The syntax of \textsc{relax} is structured natural language and boolean expressions (including standard temporal operators). It's semantics is defined in terms of temporal fuzzy logic.

\subsection{Summary of Formalisms for Robotic Systems}
\label{sec:formalismsSummary}

This section describes the formalisms currently used in the literature for specifying the behaviour of robotic systems. Table~\ref{tab:formalismSummary} summarises the formalisms found, according to the methodology described in \S\ref{sec:scope}, in use to specify the system and properties to be checked. There were more studies found that specify the system (56) than specify the properties (41). Some systems specifications are checked for deadlock or reachability, for example \cite{costelha2007modelling}; whereas some are just used for specification, such as \cite{Weyns2012}. Conversely, some of the studies only specify properties and monitor a running system~\cite{huang2014rosrv, liang2009software}, or describe a framework for dealing with property specification~\cite{whittle2009relax}. Also, the ontologies found in our literature search are all used to specify systems.

State-transition systems are the most numerous for specifying systems and temporal logics the most numerous for specifying properties. This popularity may be explained by the popularity and usability of model-checkers for these formalisms; state-transition systems are often used to describe a system's behaviour, each state of which is then checked to see if a temporal logic property holds. This is shown in Table~\ref{tab:toolSummary}, where model-checkers are the most used formal verification tool.

Despite state-transition systems and temporal logics dominating the literature found, Table~\ref{tab:toolSummary} shows that there are still a wide variety of tools being used for these formalisms. This may be because different model-checkers capture different features, such as time or probability. But this possibly presents a challenge to the sharing and extension of formal models. 

\section{Formal Verification Approaches for Robotics}
\label{sec:approaches}

This section analyses the surveyed literature in terms of the verification approaches that were used. Here, ``approach" refers to the framework(s), technique(s) or combination thereof that were used to verify that the model or implementation of the system being developed preserves the required properties; where \S\ref{sec:formalisms} discussed the \textit{types} of formalisms found in the literature, this section discusses \textit{how} these formalisms are used to verify robotic systems. If we imagine that the formalisms described in \S\ref{sec:formalisms} are different kinds of bricks, then the approaches that we describe here are the types of buildings that can be constructed from those bricks. 

We explore several distinct approaches to formal verification of robotic systems as illustrated in Table \ref{tab:approaches}. Specifically, \S\ref{sec:modelCheck} describes model-checking approaches, \S\ref{sec:theoremProve} describes approaches using theorem proving, \S\ref{sec:rvmonitor} discusses approaches using runtime verification monitors, \S\ref{sec:ifm} discusses approaches that use a combination of different formalisms (e.g. CSP and B) or approaches (e.g. model-checking and theorem proving), \S\ref{sec:frame} outlines frameworks for building verifiable robotic software; finally, \S\ref{sec:approachesSummary} summarises and answers the remaining research questions (RQ2 and RQ3).

\begin{table}
\centering
\scalebox{0.8}{
\begin{tabular}{|m{4.5cm}|m{10cm}|c|}
\hline
{\textbf{Approach}} & \textbf{References} & \textbf{Total} \\ \hline \hline

Model-Checking & 	\cite{Konur2010}, \cite{massink2013use}, \cite{konur2012analysing}, \cite{brambilla2012property}, \cite{dixon2011towards}, \cite{dixon2012towards}, \cite{iftikhar2012case}, \cite{iftikhar2014activforms}\ \cite{iglesia2015mape}, \cite{hilaire2007formal}, \cite{webster_formal_2014}, \cite{webster_toward_2016}, \cite{dantam2012robust}, \cite{filippidis2012decentralized}, \cite{webster2011formal}, \cite{proetzsch2007formal}, \cite{proetzsch2007behaviour}, \cite{chkouri2008translating}, \cite{bensalem2011formal}, \cite{basu2011rigorous}, \cite{bensalem2009designing}, \cite{abdellatif2012rigorous}, \cite{chen2012formal}, \cite{bordini2009formal}, \cite{karaman2009sampling}, \cite{fisher_verifying_2013}, \cite{Dennis2016a}, \cite{Dennis2015}, \cite{webster2011formal}, \cite{Kouvaros2017}, \cite{choi2015verification}, \cite{molnar2009system}
& 	32\\ \hline

Theorem Proving & 	\cite{mitsch2013provably}, \cite{hilaire2007formal}, \cite{kouskoulas2013certifying}	 & 	3\\ \hline

Runtime Monitoring & 	\cite{huang2014rosrv}, \cite{falcone2011runtime}, \cite{liang2009software}	 & 	3\\ \hline

Integrated Formal Methods & 	\cite{iglesia2015mape},\cite{chkouri2008translating}, \cite{akhlaki2007methodological},\cite{bonfe2012towards},  \cite{hilaire2007formal}, \cite{gudemann2008specification}, \cite{iftikhar2012case}, \cite{liang2009software}	 & 	8\\ \hline

Formal Software Frameworks /Architectures & 	\cite{falcone2011runtime}, \cite{raman2011analyzing}, \cite{bensalem2011formal}, \cite{weyns2012forms}, \cite{basu2008incremental}, \cite{basu2011rigorous}, \cite{bensalem2009designing}, \cite{proetzsch2007formal}, \cite{proetzsch2007behaviour}, \cite{Weyns2010}	 & 	10\\ \hline

%Reachability Analysis & 	\cite{hess2014formal}	 & 	1\\ \hline

%Synthesis & \cite{kress2011correct} & 1 \\\hline

%Simulation & \cite{Weyns2010}, \cite{costelha2007modelling}, \cite{gjondrekaj2012towards}, \cite{karaman2009sampling} & 4 \\\hline

\end{tabular}}
\caption{\label{tab:approaches}Summary of the approaches to formal verification found using the methodology described in \S\ref{sec:scope}. Note: some of the surveyed papers use more than one approach.}
\vspace{-1.5em}
\end{table}

\subsection{Model-Checking}
\label{sec:modelCheck}

Model-checking is the most widely used approach to verifying robotic systems; it has been used with temporal logics~\cite{Dixon2014, Izzo2016}, process algebras~\cite{Miyazawa2017}, and programs~\cite{fisher_verifying_2013, Fernandes2017}. Arguably, the popularity of model-checking owes something to the volume of publications that we found using temporal logic (Table \ref{tab:formalismSummary}), which are often used in model-checking approaches. Further, we see two main reasons for \textit{actively} choosing model-checking. First, model-checkers are automatic, which makes them relatively easy to use; second, the concept of checking every state in a model to see if a required property holds is relatively easy to explain to stakeholders without formal methods experience.

Some model-checkers can handle timed models (e.g. \Uppaal{}) or probabilistic models (e.g. PRISM). This can be very useful for dealing with robotic systems; timing constraints are often required for safe behaviour and probability can be helpful when encoding the physical environment of the robot. Both of these features can improve confidence in the results. The input to a program model-checker (such as \gls{ajpf}) is the program code itself, so checking it for properties involves \emph{symbolically executing} the code and assessing each execution against the required property~\cite{Visser2003}.

Model-checking exhaustively explores the state space, so one must be careful about the input models and properties to be checked. State explosion can be crippling to verification efforts using model-checking.  Webster et al.~\cite{webster_formal_2014} report requiring  $10^4$ MB of memory to verify each individual safety property of their  domestic assistant robot model. Similar state explosion problems are present in models of swarm robotic systems \cite{Konur2010}. Various standard approaches can mitigate state space explosion and keep models tractable~\cite{Clarke:1994:Abs,CEGAR:jnl}. For example, for swarms, symmetry reduction~\cite{Antuna2015} or abstracting the swarm to a single state machine with a counting abstraction can help~\cite{Webster2014}.

Model-checking has been used to build behaviour specifications (e.g. robot movement plans). One approach uses model-checking to build robot motion plans that satisfy properties specified in a deterministic subset of \Mucalc{}~\cite{karaman2009sampling}. A behavioural specification is built up incrementally, and checked after each expansion, until it moves the robot to its goal and satisfies the required properties. Similarly, in~\cite{guo2013revising}, a \gls{ba} representing the robot's environment is model-checked for an accepting path satisfying an \gls{ltl} specification. The accepting path is used to synthesise a motion controller, which is then revised at runtime by repeated model-checking and synthesis. Kloetzer et al.~\cite{kloetzer2011multi} use model-checking to find traces of a transition system describing the behaviour of a robot team that satisfy an \gls{ltl-x} formula. This trace is used to generate the communication and control strategy for each robot in the team. 

Clearly, model-checking is a flexible approach: it has been used for a variety of formalisms that can describe concurrent, timed, and probabilistic systems; and has been used in verification efforts for a variety of robotic systems. The increasing access to computational power and memory, combined with clever ways to reduce the state space, maintain the popularity of model-checking in the literature. The need to carefully craft a model for a particular model-checker adds to the modelling time and reduces the portability of models between tools. However, the automation and relative simplicity of the model-checking approach to verifying robotic systems mean that it remains a focus for research.

\subsection{Theorem Proving}
\label{sec:theoremProve}

Theorem proving offers the benefit of producing a formal proof of the correctness of a software system. These formal proofs can be used to provide robust evidence for certification of autonomous robotic systems. A notable example here is the use of Isabelle/HOL and temporal logic to formalise a subset of traffic rules for vehicle overtaking in Germany \cite{rizaldi2017formalising}. 
More recently, the RoboChart notation and it's associated toolset also makes use of Isabelle/HOL to verify robotic systems \cite{foster2018automating}.

Other work in this domain includes \cite{mitsch2013provably}, which uses the hybrid systems logic \gls{dl}~\cite{Platzer2007} to describe the discrete and continuous navigation behaviour of a robot rover. They then use the hybrid theorem prover, KeYmaera, to verify that the robot does not collide with stationary or moving obstacles, and maintains sufficient distance from obstacles. Mitsch et al.~\cite{Mitsch2017} update this work, verifying a less safety property that allows for imperfect sensors; adding liveness proofs, to guarantee progress; and automatically synthesising runtime monitors from the verified models, to mitigate the problems caused by the reality gap. 

In~\cite{hilaire2007formal}, OZS (a combination of Object-Z and Statecharts~\cite{Harel1987}) is used to capture the dynamic roles that an agent in a multi-robot system can play, which define the robot's interaction patterns. OZS has an operational semantics, and the STeP theorem prover is used for verification of safety and liveness properties of their model. 

Although effective, it is clear from our analysis that these kinds of approaches have not received much attention in the literature. We believe that this is a usability issue with the tools generally more difficult to master than those of other approaches.

\subsection{Runtime Monitoring}
\label{sec:rvmonitor}

Runtime monitors can be used to sidestep the problem of verifying a (needfully) complex model of a robotic system. Instead of specifying and verifying the entire system, the properties that the system has to exhibit are extracted and specified as a monitor of the system. Because the monitor is simpler than the system, it is often easier to verify. Another advantage is that runtime monitors can mitigate the problem of the reality gap (between a model and the real world) especially when used to complement offline verification. Given that a robotic system is naturally cyber-physical, and therefore malfunctions can have safety consequences, monitoring the system's behaviour at runtime can be key to ensuring safe operation~\cite{Sistla2012}.

A monitor consumes events from a system and compares them to the expected behaviour. If the system's events differ from the expectation, then it can invoke mitigating activities, including logging, flagging this to the user, or triggering behaviour to remedy the situation. Runtime monitoring cannot guarantee all system behaviours beforehand. It has received attention in the literature on runtime verification, which brings together model-checking and stream-processing technologies~\cite{RosuH05}. Examples include~\cite{falcone2011runtime, iftikhar2012case, huang2014rosrv}, and the International Conference on Runtime Verification has been running since 2006. 

Aniculaesei et al.~\cite{aniculaesei2016towards} use runtime monitors to complement design-time formal methods. The runtime monitors are built to check that design-time assumptions hold during the execution of the program. For additional confidence at design-time, they also specify the system and its physical environment using \gls{ta} with safety properties in \gls{tctl}.
Ferrando et al.~\cite{ferrando_aamas18} develop runtime verification to recognise anomalous environmental interactions and so highlight when the previous formal verification that has been carried out on an autonomous robotic system (with some environmental assumptions) becomes invalid.

Kane et al.~\cite{kane2015case} present runtime monitors for an autonomous research vehicle. These monitors are written in the $\alpha \mathcal{VSL}$ safety specification language, whose semantics is given over time-stamped traces using future-bounded, propositional \gls{mtl}. Their algorithm, \texttt{EgMon}, uses the MAUDE rewriting engine to reduce the input formulae. They extend \texttt{EgMon} with a hybrid monitoring algorithm, \texttt{HMon}, which first performs conservative checks and then as many eager checks as the remaining time permits.

Often, robotic systems are distributed; either because of the software architecture (such as \gls{ros}) or because it is a multi-robot system (such as a robot swarm). For such systems, Bauer and Falcone~\cite{bauer2012decentralised} describe a method of distributing an \gls{ltl} specification of a system's macro-level behaviour to a collection of micro-level behavioural monitors. Their approach does not assume any central information collection and ensures that the communication between monitors is minimal, but sufficient to allow the monitors to work. 

Robotic systems are intrinsically cyber-physical, combining discrete and continuous parts. The implications of this on runtime monitoring are addressed in~\cite{Sistla2012}, which describes an approach for modelling the hybrid nature of cyber-physical systems without discretising them. They use \gls{pha} to capture both the discrete and continuous elements of a system. They illustrate their approach using an electronically controlled train braking system. A similar hybrid-logic approach is taken by~\cite{Mitsch2017}, who specify the safety properties of a robot rover using \gls{dl} and then automatically synthesise runtime monitors from these verified models. 

\Rosrv{} is a runtime verification framework for robotics systems deployed on \gls{ros} \cite{huang2014rosrv} that uses a novel formal specification language for writing safety properties. Using a new configuration file, \Rosrv{} then automatically generates C++ \gls{ros} nodes to monitor the system for the specified properties. Because their approach uses normal \gls{ros} nodes and a custom configuration file, it doesn't require any changes to be made to either \gls{ros} or the application. They make the interesting observation that their approach cannot currently be verified because that ``would require a model of \gls{ros} itself'' to be able to prove that the monitors and other generated code respects the model.

Similarly, Falcone et al.~\cite{falcone2011runtime} describe \Rvbip{}, a tool that produces runtime monitors for robots in the \gls{bip} framework, which will be discussed in \S\ref{sec:frame}. They provide a formal description of the \gls{bip} component-based framework and monitors. Each monitor consumes events from the \gls{bip} system and assesses the sequence of events, producing a verdict of either true, false, \textit{currently} true or false. \Rvbip{} takes an XML description of an \gls{lts}, describing the monitor, and produces a \gls{bip} monitor for a given \gls{bip} system. Falcone et al. provide an example where they use this tool to produce a \gls{bip} system with execution order and data freshness monitors.

Liang et al.~\cite{liang2009software} describe a runtime monitoring approach that does not require any alterations or additions to the monitored system. They specify the system in Z and compare the traces from a specification animator with information from a Java debugger to check if the running program is correct, with respect to the Z specification. This has the added advantage of decoupling the monitor from the monitored program.

\subsection{Integrated Formal Methods}
\label{sec:ifm}

\Gls{ifm} refers to the integration of multiple formal methods, or a formal method with a semi- or non-formal approach, that complement each other. While still difficult, \gls{ifm} can capture several dimensions of a system at once (for example static and dynamic behaviour) for easy analysis. In our previous work, we argue that robotic systems provide the impetus for addressing the challenges of integration; furthermore, because robotic systems are inherently hybrid, they \textit{need} \gls{ifm}~\cite{Farrell2018}. An early example of this need is a study concluding that no single formalism was suitable for applying to a robot swam, and a combination of 4 formalisms would be the best approach~\cite{Hinchey2005}. 

The challenges presented in \S\ref{sec:intChallenges} are often best tackled using \gls{ifm}. However, our literature search found no generic approaches for applying \gls{ifm} to robotics. Here, we discuss some notable bespoke examples. To enable both refinement-based development and probabilistic checking, \cite{tarasyuk2012formal} uses both Event-B and a probabilistic language to check reconfigurable architectures for an on-board satellite system. The work in~\cite{Akhtar2014} combines \gls{fsp} and \gls{piadl} to capture safety and liveness properties of multi-agent robotic systems  The \gls{fsp} specifications of the relevant safety and liveness properties are transformed into \glspl{lts}, then the agent programs and architecture (described in \gls{piadl}) are checked to see if they satisfy the required properties. 

The work in~\cite{smith2014maze} combines  MAZE (an extension of Object-Z for multi-agent systems) and Back's action refinement to enable top-down development of robot swarms. In~\cite{Kamali2017}, an agent controlling a car is verified using \gls{ajpf}, and the timing properties are verified using \Uppaal{}. This is extended, in~\cite{Kamali2018} with a spatial reasoning calculus. The work verifies the cooperation between the vehicles, and the abstract behaviour of the physical vehicle. Another platooning driverless car is modelled using \gls{csp|b}~\cite{Colin2009}.

Similarly, OZS combines Object-Z~\cite{hilaire2007formal} and Statecharts~\cite{Harel1987}; it has been used to describe the state and behaviour of systems is used to formalise part of the Satisfaction-Altruism Model~\cite{Simonin2000}. This model captures the dynamic set of roles that a robot in a multi-robot team can perform. The roles define the interaction pattern of the robot. OZS has an operational semantics, which enables theorem proving. In~\cite{hilaire2007formal}, the authors describe verification of their model using the SAL model-checker and the STeP theorem prover. They also indicate an intent to provide a way to refine OZS models to code for deployment on robots.

These bespoke examples integrate two or more formal methods to obtain an approach to verification with their combined benefits. There are also examples that integrate a formalism with a non-formal notation. For example, an approach for deriving formal specifications from the real-time \gls{uml} profile UML-RT~\cite{akhlaki2007methodological}, which captures UML-RT subsystems as processes in the timed \gls{csp} derivative, CSP+T. This also provides UML-RT with a semantics in CSP+T.

RoboChart is a robotic system design notation that integrates \gls{csp} with a graphical timed state machine notation~\cite{Ribeiro2017}. The integration displayed by RoboChart provides a generic formal notation for robotic systems, which is an approach that yields a reusable and stable formal method. This reusable, stable and generic approach to \gls{ifm} is key to the success \glspl{ifm}, not just for robotic systems but more generally as well. Integrating diverse formalisms into a holistic framework remains an open problem, waiting to be solved.

In the area of automated surgical robotics the combination of \gls{uml}, \gls{ltl}, and Alloy has been used to model and reason about a number of surgical operations~\cite{bonfe2012towards}. The work employs a goal-oriented methodology where a \gls{uml} model relates goal requirements to the ``high-level structural and behavioural decomposition of the intelligent robotic system''. A number of experts (medical professionals) were interviewed and their responses were used to construct a goal model which was converted into a set of formal properties using the Alloy language where each goal is formally defined in \gls{ltl}. This model is then checked by the KodKod SAT solver. The \gls{uml} model can then be translated into executable code using the Orocos framework.

The Restore Invariant Approach (RIA) uses invariants to specify productive states and to use constraint solving techniques whenever the system is not operational to find new configurations to make the system operational again~\cite{gudemann2008specification}. This work targets self-adaptive and agent-based systems. It was used in the specification and verification of an adaptive production cell that was made up of three drill robots and two autonomous transportation units that connect them. Here, the authors make use of class diagrams with \gls{ocl} constraints, the Microsoft Robotics Studio simulator, JADEX multi-agent framework and the Alloy constraint solver.

\subsection{Frameworks for Verifiable Robotic Software}
\label{sec:frame}

Our literature search revealed a number of frameworks for developing verifiable robotic systems. These frameworks often encompass a number of the techniques already described in \S\ref{sec:approaches}, but frequently, they incorporate bespoke tools and formalisms. Although some facilitate the use of multiple verification techniques, we omit them from \S\ref{sec:ifm} since users are generally expected to apply only one of the available techniques in practice rather than integrating results from several. 

The \gls{bip} framework~\cite{Basu2006} is a toolset for modelling component-based real-time software, with a notation based on \glspl{fsm}. The basic component of a \gls{bip} model is a state transition system, labelled with C/C++ functions. These components are combined to form larger components and models can be verified using the D-Finder tool~\cite{Bensalem2011DFinder}. In~\cite{abdellatif2012rigorous}, Abdellatif et al. combine the \gls{bip} framework with the \Genom{} robot software architecture to provide a technique that can synthesise robotic control software from \gls{bip} models of the required behaviour. This enables the generation of robotic software, in \Genom{}, that is correct by construction. Abdellatif et al. show the generation of software for a wheeled rover robot, using fault injection to test that the constraints in the \gls{bip} specification are enforced by the generated software. 
 
Other work along this vein includes a metamodel-based translation from the \gls{aadl} into \gls{bip}~\cite{chkouri2008translating}. This facilitates simulation of \gls{aadl} combined with formal verification using the model-checking tools in the \gls{bip} framework, in particular, this work uses the Aldebaran model-checker. Furthermore, Falcone et al.~\cite{falcone2011runtime} integrate runtime verification into the \gls{bip} framework. They describe their approach and a tool, \Rvbip, which generates monitors for \gls{bip} to check specifications at runtime. Related to this are attempts to integrate \gls{bip} and the LAAS architecture which is based on \Genom{}~\cite{basu2008incremental, bensalem2009designing}. 

The Averest framework provides ``tools for verifying temporal properties of synchronous programs'' that are written in the Quartz language~\cite{proetzsch2007behaviour}. The Quartz programs are translated into Averest Interchange Format (AIF) and then verified against \gls{ltl} specifications using the Beryl model-checker or other third party tools that have interfaces to them such as SMV. Furthermore, the framework has a code generation tool called Topaz that can output Verilog, VHDL and C code.

Chen et al.~\cite{chen2012formal, chen2013formal} propose a computational framework for automatic synthesis of control and communication strategies for robot teams, using task specifications that are written as regular expressions. Their objective is to ``generate provably correct individual control and communication strategies'' from regular expressions, which are used to create \gls{fsa}. They demonstrate their approach using a \gls{rule}, where autonomous cars must navigate and avoid collisions. Their technique builds on \gls{ltl} model-checking and uses JFLAP and MatLab~\cite{chen2013formal}.

The \gls{forms} provides a reference model that can guide developers as to how to formulate a Z specification of a self-adaptive system~\cite{weyns2012forms}. This is built upon MAPE-K and supports agents and formal refinement of specifications in Z.

Kim et al.~\cite{kim_formal_2005} use the Esterel framework to verify the stopping behaviour of a home service robot. This framework offers facilities for specifying and verifying robotic systems using model-checking. Esterel can also be translated into an associated C or C++ program and the semantics of an Esterel program is given in terms of the \gls{fsm} that it describes. To enable systems to be checked for temporal logic properties, the authors translate them into observers in Esterel.

\subsection{Summary of Approaches}
\label{sec:approachesSummary}

This section discussed the formal verification approaches in the literature as summarised in Table~\ref{tab:approaches}. The most popular approach is model-checking, which is reflected in the dominance of temporal logics and state-transition systems for the specification of robotic systems (\S\ref{sec:formalisms}, Table~\ref{tab:formalismSummary}). We speculate that the prevalence of model-checking in the literature may be because it is easy for developers to understand and trust that do not have a formal methods background since  it is automatic and conceptually similar to exhaustive testing. Frameworks for verifiable robotic systems were the next most popular approach and most of these incorporated a bespoke model-checker. However, it is not clear, in practice, just how effective these inbuilt verification tools are.

As outlined earlier, \gls{ifm} are necessary in the verification of  robotic systems due to their size and complexity. This is an active area of research with tools still under development. We anticipate that, in the future, this approach will feature more in the verification of autonomous robotic systems. The least popular approaches were theorem proving and runtime monitoring. In general, theorem provers require specialist knowledge to use and we believe that this is why it is currently not as popular as other approaches. Although effective in the running system, runtime monitoring is often seen as a separate step in the verification process after static verification has taken place and provides little in the way of design-time benefits.

\section{Conclusion}
\label{sec:summaryAndObs}
\label{sec:discussion}

\S\ref{sec:relatedWork} discussed related surveys and differentiated them from our work. Our work incorporates the topics of formal methods for self-adaptive systems (surveyed in~\cite{Weyns2012}) and swarms (from~\cite{rouff2004formal, Rouff2004}). We also identify work from several categories that a survey on safety-critical robotics~\cite{Guiochet2017} lists as areas to focus on. Specifically, these are modelling of a robot's physical environment, formal verification of robotic systems, correct-by-construction controller synthesis, and being able to provide certification evidence. This last point is particularly important, and is a key focus in the current literature. Finally, the work that we have surveyed is squarely within the category of verifying applications for internal correctness; one of three main challenges for the automatic verification of autonomous systems~\cite{simmons_towards_2000}. Some of the current literature focusses on the external correctness of the software, for example how it will safely interact with a robotic software framework like \gls{ros} or \Genom{}. However, the suggestions that special purpose language interpreters or compilers must be verified has not been borne out, as many robotic applications are written in general purpose languages such as Python, Java, or C++. This, of course, leaves the tools for these languages to be verified, a larger challenge due to the generality of the languages.

\S\ref{sec:scope} outlined the research questions that this paper investigates. In this section, we use the results of our survey to derive potential answers to these research questions. The answers also provide the structure for a discussion of our observations of the current state of the literature.\\

\noindent\textbf{RQ1}, asked what the challenges are when formally specifying and verifying (autonomous) robotic systems. We identified two types of challenge: those \textit{external} to the robotic system, discussed in \S\ref{sec:challenges}; and those \textit{internal} to the robotic system, discussed in \S\ref{sec:autochall}. The external challenges that we identified are: modelling the robotic system's external environment, and providing sufficient evidence for public trust and certification. The internal challenges that we identified are the use of: agent-based, multi-robot, and adaptive/reconfigurable systems. Reconfigurability is key to safely deploying robots in hazardous environments and vastly more work needs to materialise in order to ensure the safety of reconfigurable autonomous systems. Therefore, we see a clear link between a robotic system reacting to the changes in its external environment, and reconfigurable systems. 

Similarly, \textit{rational} agent-based systems that can explain their reasoning provide a good route for providing evidence for public trust or certification bodies. This is because they enable the crucial transparency that trust and certification needs. An example of the transparency required is the \textit{ethical black box} proposed by~\cite{Winfield2017}, which records the sensor input and internal state of the system. A rational agent can provide reasons for its choices, based in the input and internal state information.\\

\noindent\textbf{RQ2}, asked what are the current formal methods used for tackling the challenges identified by answering \textbf{RQ1}. Our analysis, presented in \S\ref{sec:formalisms} and \S\ref{sec:approaches}, shows that temporal logics, state-transition systems, and model-checkers are the most prominent formalisms and approach in the literature over the past decade. We speculate that this is due to the fact that temporal logics and state-transition systems allow abstract specification, which is useful earlier in the development process, and because model-checking as an approach is generally easy to explain to stakeholders who do not have experience using formal methods.\\

\noindent\textbf{RQ3}, asked what the limitations are of the best practice formalisms and approaches to verification that were identified in the answer to \textbf{RQ2}. One limitation appears to be a resistance to adopting formal methods in robotic systems development~\cite{Lopes2016}. The perception is that applying formal methods is a complicated additional step in the engineering process, which prolongs the development process while not adding to the value of the final product. A lack of appropriate tools also often impedes the application of formal methods. These are long-standing opinions, for example the authors of~\cite{Knight1997}, which was published 19 years prior to~\cite{Lopes2016}. There are, however, notable examples of industrial uses of formal methods, such as those examined in~\cite{Woodcock2009}.

The wide variety of tools for the same formalism (indicated in Table \ref{tab:toolSummary}) also points to a problem: the lack of interoperability between formalisms. Often, models or specifications of similar components are incompatible and locked into a particular tool. This suggests that a common framework for translating between, relating, or integrating different formalisms, would prove useful in smoothing the conversion between formalisms or tools. Further, this would serve a growing need to capture the behaviour of complex systems using a heterogeneous set of formalisms, each suited to the component being modelled or the properties of interest. This is currently an open problem in formal methods for robotic systems.

As mentioned in \S\ref{sec:extChallenges}, there is a lack of clear guidance when it comes to choosing a suitable formal method for a particular system. With respect to autonomus robotic systems, Table~\ref{table:littable} lists the formalisms/tools used and the corresponding case study that was presented in the surveyed literature. This provides some guidance when choosing an appropriate formal method for a given system but we leave a more detailed analysis of this as future work.

Another limitation faced by formal methods for robotic systems (and more generally) is that of formalising the \textit{last link}, the step between a formal model and program code. We found few examples of formal methods for robotics that also produced runnable program code from a specification, and, in these examples, it was often unclear as to how the code relates back to the original specification model. Such last link translation steps require a formal underpinning to ensure that the model is represented by the code. The lack of clarity about this limitation points to another: a lack of open sharing of models, code, and realistic case studies that are not tuned for a particular formalism. 

Field tests and experiments using simulations are both useful tools for robotic systems development; but formal verification is crucial, especially at the early stages of development when field tests of the control software are infeasible (or dangerous). A focussed research effort on the combination or integration of formal methods should improve their use in robotic systems development, because no one formalism is capable of adequately capturing that all aspects of a robotic system behave as expected. Ensuring that these tools are usable by developers and providing similar features in an \gls{ide} would also improve their uptake by simplifying their use. Work in this area could lead to an Integrated \textit{Verification} Environment, allowing the use of different formalisms using same developer front-end, connecting them to their respective tools, and providing helpful \gls{ide}-like support. 

We have identified a number of \textit{threats to validity} for this work, and we have taken measures to mitigate them where possible. In particular, researcher bias was minimised by explicitly stating the scope and research questions at the beginning of the review and having two researchers analyse the search results in tandem. Restricting our search to those papers published within the last ten years may affect the completeness of our search results, however, by `snowballing' we have explored and discussed papers that did not fall within this range throughout the narrative in this report. Furthermore, our use of Google Scholar limits our search to those papers that were found using Google's algorithms. Again, our use of ``snowballing'' helps to mitigate this threat. 

This paper has analysed the current state-of-the art literature particular to the formal specification and verification of autonomous robotic systems. To this end, this survey identifies future directions in this field.

\bibliographystyle{abbrv}
\bibliography{bibliography,search}

\end{document}